\documentclass[journal]{IEEEtran}

\usepackage{graphicx}
\usepackage{times}
\usepackage{soul}
\usepackage{url}
\usepackage[colorlinks,citecolor=blue]{hyperref}
\usepackage[utf8]{inputenc}
\usepackage[small]{caption}
\usepackage{graphicx}
\usepackage{amsmath}
\usepackage{amsthm}
\usepackage{amssymb}
\usepackage{booktabs}
\usepackage{algorithm}
\usepackage{algpseudocode}
\usepackage{xcolor}
\usepackage{colortbl}
\usepackage{algorithm}
\usepackage{algpseudocode}

\usepackage{pifont}

\newcommand{\review}[1]{{\textcolor{black}{#1}}}
\usepackage{multirow}
\newcommand{\tabincell}[2]{\begin{tabular}{@{}#1@{}}#2\end{tabular}}

\newcommand{\best}[1]{\textbf{\textcolor{red}{#1}}}
\newcommand{\secondbest}[1]{\textbf{\textit{{\textcolor{violet}{#1}}}}}
\definecolor{LightCyan}{rgb}{0.88,1,1}
\definecolor{LightBlue}{rgb}{0.4588,0.8235,1}
\definecolor{TableColor}{rgb}{0,0,0}

\makeatletter
\newcommand*{\@rowstyle}{}
\newcommand*{\rowstyle}[1]{
	\gdef\@rowstyle{#1}%
	\@rowstyle\ignorespaces%
}
\newcolumntype{=}{
    >{\gdef\@rowstyle{}}%
}
\newcolumntype{+}{
    >{\@rowstyle}%
}
\newcolumntype{C}[1]{>{\centering\arraybackslash}p{#1}}
\makeatother

\title{Robust AI-Synthesized Speech Detection Using Feature Decomposition Learning and Synthesizer Feature Augmentation}

\author{Kuiyuan~Zhang,
        Zhongyun~Hua,
        Yushu~Zhang,
        Yifang~Guo,
        and Tao~Xiang
\thanks{This work was supported by the National Key R\&D Program of China under Grant 2022YFB3103500 and by the National Natural Science Foundation of China under Grant 62071142.}
\thanks{Kuiyuan~Zhang and Zhongyun~Hua are with School of Computer Science and Technology, Harbin Institute of Technology, Shenzhen, Guangdong 518055, China (e-mail: zkyhitsz@gmail.com; huazyum@gmail.com).}
\thanks{Yushu Zhang is with the College of Computer Science and Technology, Nanjing University of Aeronautics and Astronautics, Nanjing, Jiangsu 210016, China (e-mail: yushu@nuaa.edu.cn).}
\thanks{Yifang~Guo is with Alibaba Group, Hangzhou, Zhejiang 310056, China (e-mail: guoyifang@gmail.com).}
\thanks{Tao Xiang is with the College of Computer Science, Chongqing University, Chongqing 400044, China (e-mail: txiang@cqu.edu.cn).}
\thanks{This work has been submitted to the IEEE for possible publication. Copyright may be transferred without notice, after which this version may no longer be accessible.}
}

\begin{document}

\maketitle

\begin{abstract}
AI-synthesized speech, also known as deepfake speech, has recently raised significant concerns due to the rapid advancement of speech synthesis and speech conversion techniques. Previous works often rely on distinguishing synthesizer artifacts to identify deepfake speech.
However, excessive reliance on these specific synthesizer artifacts may result in unsatisfactory performance when addressing \review{speech signals} created by unseen synthesizers. In this paper, we propose a robust deepfake speech detection method that employs feature decomposition to learn synthesizer-independent content features as complementary for detection. Specifically, we propose a dual-stream feature decomposition learning strategy that decomposes the learned speech representation using a synthesizer stream and a content stream. The synthesizer stream specializes in learning synthesizer features through supervised training with synthesizer labels. Meanwhile, the content stream focuses on learning synthesizer-independent content features, enabled by a pseudo-labeling-based supervised learning method. This method randomly transforms speech to generate speed and compression labels for training. Additionally, we employ an adversarial learning technique to reduce the synthesizer-related components in the content stream. The final classification is determined by concatenating the synthesizer and content features. To enhance the model's robustness to different synthesizer characteristics, we further propose a synthesizer feature augmentation strategy that randomly blends the characteristic styles within real and fake audio features and randomly shuffles the synthesizer features with the content features. This strategy effectively enhances the feature diversity and simulates more feature combinations. Experimental results on three deepfake speech benchmark datasets demonstrate that our model achieves the state-of-the-art robust detection performance across various evaluation scenarios, including cross-method, cross-dataset, and cross-language evaluations.

\end{abstract}

\section{Introduction}

With the rapid advancement of deep learning techniques, deepfake technology, including the synthesis and manipulation of multimedia content, has become increasingly accessible~\cite{Siwei2020-deepfake_review}. The recent advancements in deepfake generation methods have enabled the creation of multimedia content with \review{remarkable} reality, presenting a significant threat to the security of multimedia information~\cite{Aynur2022-deepfake_review}, such as impersonation attack~\cite{franceschi-bicchierai2020listen}, reputation damage, or online harassment~\cite{burgesstelegram2020}. Despite the considerable focus on deepfake video detection, research on deepfake speech detection remains relatively underdeveloped~\cite{frank2021wavefake}.

Deepfake speech, also known as AI-synthesized speech, involves the synthesis or manipulation of speech waveforms to replace the original audio content with artificially generated content. Two common deepfake speech generation methods are text-to-speech (TTS) and voice conversion (VC), both of which typically utilize neural vocoders to produce audio waveforms based on temporal-frequency representations. TTS methods allow for the synthesis of audio with specific voice styles from text inputs~\cite{kim2024P-Flow-TTS}, while VC methods enable the modification of voice styles while retaining the original content~\cite{shan2024phoneme-One-shot-VC}. The advancement of both TTS and VC technologies has significantly increased the challenge of distinguishing between genuine and fake \review{speech signals} using human perception~\cite{almutairi2022review}.

To address the potential threat caused by deepfake speech, it is imperative to develop effective detection methods capable of distinguishing between genuine and fake \review{speech signals}~\cite{sunAISynthesizedVoiceDetection2023-LibraSeVoc}. Initially, early deepfake speech detection methods primarily relied on specific statistical features inherent to audio signals, such as Mel-frequency cepstral coefficient (MFCC)~\cite{wang2017spoofing}, linear frequency cepstral coefficients (LFCC)~\cite{ding2021robustness}, constant Q cepstral coefficients (CQCC)~\cite{zhan2022-DetectingSpoofedSPeeches-CQCC}, and Fourier bi-spectrum~\cite{albadawy2019detecting}. However, these methods have shown limited effectiveness against the rapid development of deepfake speech generation techniques.

Recently, some well-designed deep learning models have emerged to address the challenge of deepfake speech detection. These models include multi-task learning networks~\cite{sunAISynthesizedVoiceDetection2023-LibraSeVoc}, unsupervised pre-training models~\cite{lv2022fakeAudioDetection}, graph neural networks~\cite{jung2022aasist}, multi-view-based networks~\cite{yang2024robustADD-multi-view}, and ResNet-based networks~\cite{hua2021towards-TSSDNet}. These models directly learn discriminative features from speech and perform well in intra-dataset evaluation. However, they exhibit unsatisfactory performance on unseen synthesizers or real-word data~\cite{muller2022does}. This is attributed to the inherent limitations of their feature learning strategies, which cause the detection model to focus on specific synthesizer artifacts overly. Consequently, these methods are ineffective when dealing with new types of synthesizers. 


In this study, we propose a new approach for robust deepfake speech detection using feature decomposition learning and synthesizer feature augmentation. Our goal is to enhance detection robustness by learning synthesizer-independent content features as complementary features. We first design a dual-stream feature decomposition learning strategy that employs a synthesizer stream and a content stream to decompose the speech representation learned from the backbone model. The synthesizer stream is responsible for learning the synthesizer-related features through supervised training with synthesizer labels, while the content stream focuses on learning synthesizer-independent content features. As direct content-related labels for training are unavailable, we employ a pseudo-labeling-based supervised learning method for the content stream. This method generates compression and speed labels for training by randomly altering speech characteristics through applying various compression levels and codecs, as well as adjusting the speech speed. Additionally, we employ an adversarial learning method to reduce the synthesizer-related components in the content stream. This involves integrating an adversarial loss to force the classification probabilities of synthesizers based on content features to resemble random guessing. For classification, we concatenate the synthesizer and content features to determine whether the input speech is synthesized. To further enhance the detection robustness of our method on different synthesizer characteristics, we propose a feature augmentation strategy consisting of feature blending and shuffle operations. The feature blending operation randomly merges the characteristic styles within each class of feature to enhance feature diversity, while the feature shuffle operation mixes synthesizer features with content features to simulate more synthesizer-content feature combinations. 
The main contributions of this work are summarized as follows:
\begin{itemize}
    \item  We develop a robust detection model that employs dual-stream feature decomposition learning to detect AI-synthesized speech. Different from previous methods overly relying on specific vocoder artifacts, our method employs feature decomposition to learn vocoder-independent features as the complementary feature for detection.

    
    \item We propose a synthesizer feature augmentation strategy to enhance the model's robustness to different synthesizer characteristics and synthesizer-content feature combinations.
    
    \item We conduct extensive experiments on three benchmark datasets, and the results demonstrate that our method achieves state-of-the-art detection performance and exhibits robust generalizability across diverse synthesizer methods, datasets, and languages.
\end{itemize}

This paper is structured as follows: a literature review is presented in Section~\ref{sec:related_work}. The architecture and methodology of our two-stream network are presented in Section~\ref{sec:method}.
Section~\ref{sec:experiments} presents the implementation details and the experimental results. Sections~\ref{sec:ablation} illustrates ablation studies and discusses the effectiveness of model components. Finally, Section~\ref{sec:conclusion} summarizes the conclusion and future work.

\begin{figure*}[t]
    \centering
    \includegraphics[width=0.99\textwidth]{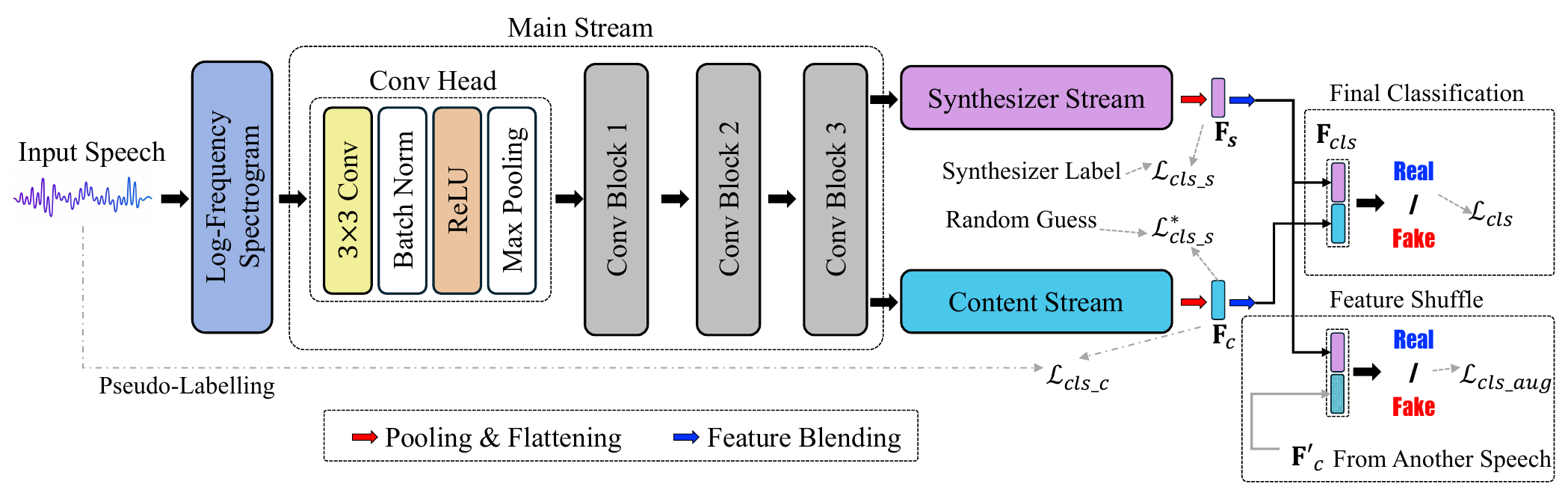}
    \caption{Network architecture of our method. A main stream is used to learn robust speech representation from the log-scale frequency spectrogram of the input speech. Subsequently, a dual-stream learning strategy, comprising a synthesizer stream and a content stream, is employed to decompose the learned speech representation. The final classification is performed based on the concatenation of the synthesizer and content features. A synthesizer feature augmentation strategy consisting of feature blending and feature shuffle operations is employed to enhance the model's robustness to different synthesizer characteristics and synthesizer-content feature combinations.}
    \label{fig:model}
\end{figure*}

\section{Related Works}
\label{sec:related_work}
\subsection{Deepfake Speech Generation}
Speech synthesis has a long time history and finds widespread application in various domains such as speech assistants, urban transportation systems, and aiding individuals with disabilities. The early-stage TTS and VC methods have limited capability, making the synthetic speech easily distinguishable to the human ear. However, advancements in TTS and VC technologies have brought synthetic speech increasingly closer to real human speech, thereby posing a growing threat to information security.

TTS methods~\cite{jiang2023sedeptts} synthesize audio waveform directly from the input text, whereas VC methods~\cite{wang2022drvc} take existing audio as input and modify voice characteristics, such as timbre and pitch, to simulate a different speaker. Despite their different inputs, many TTS and VC methods have a same final processing step: utilizing vocoders to synthesize output audio waveforms from audio spectrograms. Existing neural vocoders can be roughly divided into auto-regressive models (e.g., WaveNet~\cite{oord2016wavenet} and WaveRNN~\cite{kalchbrenner2018efficient-wavernn}), diffusion models (e.g., Guided-TTS~\cite{kim2022guided-tts} and Grad-TTS~\cite{popov2021grad-tts}), and GAN-based models (e.g., MelGAN~\cite{kumar2019melgan} and HiFi-GAN~\cite{song2023dspgan}). Auto-regressive models generate audio signals sequentially by conditioning on previously generated audio samples. Diffusion models learn a continuous diffusion process to produce realistic and diverse audio samples. GAN-based models leverage adversarial training to produce high-quality audio samples. 

In addition to vocoder components, TTS methods usually incorporate other components, such as Mel-spectrogram generators, which convert input text into Mel-spectrogram. In 2021, Kim $et~al.$~\cite{kim2021conditional-VITS} designed a groundbreaking end-to-end TTS model known as VITS, which directly synthesizes speech waveforms from input text without the need for intermediate spectrogram generation. The VITS model utilizes a variational autoencoder (VAE) network and is trained using adversarial learning techniques. This pioneering work has inspired numerous advancements in both TTS~\cite{NEURIPS2022_HierSpeech} and VC~\cite{lei2023_UniSyn} methods based on VAEs architectures.

In addition to these existing methods, research on TTS and VC methods is still progressing intensely. Therefore, it is crucial to develop robust deepfake speech detection methods to address potential threats by emerging synthesizer types.

\subsection{Deepfake Speech Detection}
\label{sec:related_work_deepfake}
In the early stage, speech detection primarily focused on tasks such as audio authentication for speech synthesis~\cite{wenger2021hello}, replay attack detection~\cite{lavrentyeva2017audio-replay-attack}, and speaker verification~\cite{jung2020improved-RawNet2}. To address the challenge of detecting deepfake speech, researchers have turned to classical speech detection methods for detecting AI-generated audio. For instance, Frank $et~al.$~\cite{frank2021wavefake} proposed a new deepfake speech dataset employing various vocoders, using classical models like RawNet2~\cite{jung2020improved-RawNet2} and LFCC-GMM as detection baselines. Experimental results on this dataset demonstrate the effectiveness of these models across certain vocoder methods. Moreover, the \textit{deepfake} database in the ASVspoof 2021 Challenge~\cite{ASVspoof2021} incorporates more than 100 vocoders and employs CQCC-GMM, LFCC-GMM, LFCC-LCNN~\cite{lavrentyeva2019stc-LCNN}, and RawNet2 as baselines.

Recently, various methods~\cite{xie2024domain_ASDG,sunAISynthesizedVoiceDetection2023-LibraSeVoc,jung2022aasist} for deepfake speech detection have been explored. 
Sun $et~al.$~\cite{sunAISynthesizedVoiceDetection2023-LibraSeVoc} introduced a multi-task learning approach, which utilizes the RawNet2 model as the backbone and incorporates an extra classifier to identify the vocoder. We abbreviate it as RawNet2-Voc in the following sections. Jung $et~al.$~\cite{jung2022aasist} developed AASIST, an audio anti-spoofing system based on graph neural networks. It incorporates a novel heterogeneous stacking graph attention layer. Additionally, Lv $et~al.$~\cite{lv2022fakeAudioDetection} and Wang $et~al.$~\cite{Wang2022FullyE2E} integrated unsupervised pre-training models to construct robust fake audio detection systems. Furthermore, classical vision architectures, such as Transformer~\cite{Ulutas-Transformer2023} and ResNet~\cite{hua2021towards-TSSDNet}, have also been employed in deepfake speech detection.

\review{
To further improve the detection generalization, some methods have utilized side-task (multi-task) learning to enable the model to learn useful features and representations that enhance deepfake speech detection~\cite{cuccovillo_audio_2023_f0,cuccovillo2024audio_f0f1f2}. For example, several researchers used Transformers to encode input spectrograms and predict the fundamental frequency (f0) trajectory~\cite{cuccovillo_audio_2023_f0} or the trajectory of the first phonetic formants~\cite{cuccovillo2024audio_f0f1f2} as side tasks to aid in learning speech representation.
Additionally, adversarial and contrastive learning strategies have been employed to develop more robust speech representation and ideal feature spaces~\cite{xie2024domain_ASDG,wu2024CLAD,goel23_Attention_Contrastive}. For example, the method in~\cite{xie2024domain_ASDG} utilizes adversarial learning and contrastive loss to create a feature space that aggregates real speech and separates fake speech from different domains, thereby improving generalizability. The authors in~\cite{wu2024CLAD} proposed a contrastive learning-based detector to enhance robustness against manipulation attacks by incorporating contrastive learning and a length loss to minimize variations caused by manipulations and cluster real speech signals more closely in the feature space. 
}

Despite the satisfactory performance of these existing methods on their evaluation datasets, their detection capabilities are limited when applied to real-world data, as highlighted by the study in~\cite{muller2022does}. This limitation stems from their inherent reliance on specific synthesizer artifacts, rendering them overly dependent on characteristics that may not be present in new types of synthesizers. Therefore, it is crucial to develop robust deepfake speech detection methods that can learn synthesizer-independent features.

\review{In this paper, we propose feature decomposition learning, which differs from traditional multi-task learning. While multi-task learning handles separate tasks in parallel and optimizes for multiple objectives~\cite{cuccovillo_audio_2023_f0,cuccovillo2024audio_f0f1f2}, our method focuses on a single detection task but decomposes the feature space into different streams to learn distinct representations within the same task. 
Previous adversarial and contrastive learning-based methods~\cite{xie2024domain_ASDG,wu2024CLAD,goel23_Attention_Contrastive} concentrate on clustering feature spaces from different domain samples. However, their detection robustness is often limited by the restricted variety of domains and synthesizers used for training. In contrast, our method emphasizes feature decomposition and synthesizer independence to enhance generalizability, significantly improving the robustness of our detection approach. }

\section{Proposed Method}
\label{sec:method}

This section describes the network structure and each component of our method. We first describe the overview pipeline and then introduce the details of the proposed modules.

To illustrate the feature learning and speech classification processes, we denote the input speech as $\mathbf{X} \in \mathbb{R}^{C \times L}$, where $C$ represents the number of audio channels and $L$ indicates the length of the waveform. Typically, the number of channels $C$ is either 1 or 2.


\subsection{Overview Pipeline}

The overall structure of our method is illustrated in Fig.~\ref{fig:model}. Given the input audio $\mathbf{X}$, our method extracts its log-scale frequency spectrogram $\mathbf{F} \in \mathbb{R}^{C \times H \times W}$ as the base feature for classification. The extraction of the log-scale spectrogram is calculated as follows:
\begin{equation}
    \begin{aligned}
        \mathbf{F} = \operatorname{log}_{e}(\operatorname{STFT}(\mathbf{X}) + 1e^{-7}).
    \end{aligned}
\end{equation}
Here, STFT denotes the Short-Time Fourier Transform for spectrogram extraction. We set the window size as 512 and the hop length as 187 in STFT. The obtained log-scale frequency spectrogram $\mathbf{F}$ is of shape (257, 257) in the spatial dimension.

Our main stream takes the log-scale spectrogram $\mathbf{F}$ as input and utilizes a convolutional head along with three convolutional blocks to learn robust speech representation. In each block, the number of feature channels in the learned feature maps increases while the spatial dimensions are down-sampled. This reduces computation complexity and allows for the learning of more abstract speech representation. Subsequently, we construct a dual-stream architecture comprising a content stream and a synthesizer stream. This architecture aims to decompose the learned audio representation into synthesizer features and synthesizer-independent content features, respectively.

Finally, we fuse the synthesizer and content features to classify whether the input speech is real or fake. Additionally, we employ a synthesizer feature augmentation strategy to enhance detection robustness.

\subsection{Main Stream}

We utilize the ResNet~\cite{he2016deep-resnet}, a classical convolutional neural network, as the backbone of our main stream to learn robust speech representation. ResNet utilizes residual connections between successive convolutional layers to address the issue of gradient degradation. It has shown effectiveness in various classifications and localization tasks. \review{The ResNet architecture primarily consists of basic residual blocks and convolutional blocks.} The basic residual block comprises $3\times 3$ convolutional layers, batch normalization layer~\cite{santurkar2018does-BN}, and ReLU activation. The convolutional block is implemented by adding a convolutional layer with strides of two and stacking the basic residual blocks.

It should be noted that we employ a lightweight version of ResNet, ResNet18, as the backbone in our method. ResNet18 has four convolutional blocks, and each convolutional block contains two basic residual blocks. The full implementation code of ResNet18 can be found on Github\footnote{https://github.com/hche11/VGGSound/blob/master/models/resnet.py}.
In our main stream, we employ the first three convolutional blocks of the ResNet18 architecture as the backbone. The fourth convolutional block serves as the final feature extractor in both the synthesizer stream and content stream.

\subsection{Synthesizer Stream}
We design the synthesizer stream to learn specific synthesizer features. Taking the speech representation $\mathbf{F}_{H}$ acquired by the main stream as input, the synthesizer stream \review{$\operatorname{S_{syn}}$} employs a convolutional block followed by a 2D average pooling layer to learn synthesizer features $\mathbf{F}_s \in \mathbb{R}^{N}$.

To ensure that the synthesizer stream learns specific synthesizer features, we impose a supervised learning task on it. Specifically, we require it to predict the labels of the synthesizer method. Assume that there are $N_s$ synthesizer methods in the deepfake speech dataset, and $y_s \in [0, N_s]$ denotes the index of the synthesizer. Note that $y_s=0$ indicates real speech. The prediction process for the synthesizer labels is illustrated as follows:
\begin{equation}
    \begin{aligned}
    \mathbf{F}_s &= \operatorname{Pool}_{avg}(\review{\operatorname{S_{syn}}}(\mathbf{F}_{H})), \\
    \hat{y}_s &=  \operatorname{softmax}(\review{\operatorname{L_s}}(\mathbf{F}_s)),
    \end{aligned}
\end{equation}
where $\operatorname{Pool}_{avg}$ denotes the 2D average pooling layer, \review{$\operatorname{L_s}$} represents a linear classifier with weights size $(N_s+1)\times N$, $\hat{y}_s \in \mathbb{R}^{N_s+1}$ denotes the \review{logits}.

The classification loss for the synthesizer stream is computed as follows:
\begin{equation}
    \mathcal{L}_{cls\_s} = \operatorname{CE}(\hat{y}_s,  {y}_s),
\end{equation}
where CE denotes the cross-entropy loss, which is commonly used for multi-class classification.

In addition to the classification loss, we incorporate a contrastive loss to enhance similarity for samples with the same synthesizer labels and decrease similarity for samples with different synthesizer labels~\cite{cai2022you-LAVDF}. Assume that $\mathbf{Z}=\{\mathbf{z}^1,\mathbf{z}^2, \cdots, \mathbf{z}^B \}$ is a batch of features and $\mathbf{y} = \{y^1, y^2, \cdots, y^B\}$ are the corresponding labels, where $B$ denotes the batch size. The contrastive loss is defined as follows:
\begin{equation}
    \begin{aligned}
        CL(\mathbf{Z}, \mathbf{y})= & \frac{1}{B^2} \sum_i^B\left(\sum_{j: y^i=y^j}^B\left(1-s(\mathbf{z}^i, \mathbf{z}^j)\right)\right. \\
        & \left.+\sum_{j: y^i\neq y^j}^B \max \left(s(\mathbf{z}^i, \mathbf{z}^j)-\alpha, 0\right)\right),
    \end{aligned}
\end{equation}
where $s(\mathbf{z}^i, \mathbf{z}^j)=\frac{\mathbf{z}_i \cdot \mathbf{z}_j}{\left\|\mathbf{z}_i\right\|\left\|\mathbf{z}_j\right\|}$ represents the cosine similarity function, and $\alpha$ is the margin parameter to control the similarity for label-unmatched sample pairs. The contrastive loss for synthesizer features is calculated as follows:
\begin{equation}
    \begin{aligned}
        \mathcal{L}_{con\_s} &= CL(\{\mathbf{F}_s\}_0^B, \{y_s\}_0^B).\\
    \end{aligned}
\end{equation}

By training the synthesizer stream with the losses $\mathcal{L}_{cls\_s}$ and $\mathcal{L}_{con\_c}$, the synthesizer stream can effectively learn discriminative synthesizer features.

\subsection{Content Stream}
We design the content stream to learn synthesizer-independent content features. Over-reliance on synthesizer features may lead to poor generalizability on \review{speech signals} synthesized by unseen synthesizers. Therefore, we aim to utilize the synthesizer-independent content features as complementary of synthesizer features for final classification.

Similar to the synthesizer stream, the content stream \review{$\operatorname{S_{content}}$} consists of a convolutional block followed by a 2D average pooling layer to learn content features $\mathbf{F}_c \in \mathbb{R}^{N}$ from the hidden states $\mathbf{F}_{H}$. This process is illustrated as follows:
\begin{equation}
    \mathbf{F}_c = \operatorname{Pool}_{avg}(\review{\operatorname{S_{content}}}(\mathbf{F}_{H})).
\end{equation}
To ensure that $\mathbf{F}_c$ represents synthesizer-independent content features, we design the loss function of the content stream based on pseudo-labeling-based supervised learning and adversarial learning.

\subsubsection{Pseudo-labeling-based Supervised Learning}

To ensure that the model learns content-oriented speech features, we must define suitable optimization goals. We do not take deepfake detection as the supervised task since it will inevitably cause the model to learn synthesizer-related features. Besides, an ideal task should be dataset-agnostic, thus enabling models to be trained on different datasets without manually labeling the data. To this end, we propose a pseudo-labeling-based supervised learning method, which obtains pseudo-label by speech transformation and can be applied to different datasets. Specifically, we randomly change the speech speed to generate the speed label and randomly compress the speech using different codecs and bit-rates to generate the compression label. Both of these pseudo labels can be used for supervised training, enabling the content stream to learn synthesizer-independent features. \review{Note that changing the speed of a speech signal is equivalent to resampling it at a different sampling rate. We achieve this by resampling the signal using Sinc interpolation with a Kaiser window to adjust the speed.}

We denote that the speech transformation has $N_1$ compression and $N_2$ speed settings. We define ${y}_c^1\in \mathbb{R}^{N_1}$ and  ${y}_c^2\in \mathbb{R}^{N_2}$ as the compression and speed labels, respectively. The prediction process of the compression and speed labels is illustrated as follows:
\begin{equation}
    \begin{aligned}
    \hat{y}_c^1 &=  \operatorname{softmax}(\review{\operatorname{L_c^1}}(\mathbf{F}_c)),\\
    \hat{y}_c^2 &=  \operatorname{softmax}(\review{\operatorname{L_c^2}}(\mathbf{F}_c)),
    \end{aligned}
\end{equation}
where \review{$\operatorname{L_c^1}$} is a linear classifier with weights size of $N_1 \times N$, \review{$\operatorname{L_c^2}$} is a linear classifier with weights size of $N_2 \times N$,  $\hat{y}_c^1 \in \mathbb{R}^{N_1}$ and $\hat{y}_c^2 \in \mathbb{R}^{N_2}$ denote the \review{logits} for the compression prediction task and the speed prediction task, respectively. Subsequently, the classification loss for the content stream is computed  as follows:
\begin{equation}
    \mathcal{L}_{cls\_c} = \operatorname{CE}(\hat{y}_c^1,  {y}_c^1) + \operatorname{CE}(\hat{y}_c^2,  {y}_c^2).
\end{equation}

\subsubsection{Adversarial Learning}

To force the content stream to learn only synthesizer-independent features, we employ adversarial learning to suppress its accuracy in the prediction of synthesizer methods. Concretely, we add an adversarial loss as the objective loss, which is defined as:
\begin{equation}
    \begin{aligned}
        \hat{y}_s^* &=  \operatorname{softmax}(\review{\operatorname{L_s}}(\mathbf{F}_c))\\
        \mathcal{L}_{cls\_s}^* &= \operatorname{CE}(\hat{y}_s^*,  y(N_s+1)),
    \end{aligned}
\end{equation}
where $y(N_s+1)$ denotes a vector of length $N_s+1$ with all values being $\frac{1}{N_s+1}$. By reducing this adversarial loss, the prediction of the synthesizer methods based on content features becomes random guessing. In other words, this can reduce the synthesizer-related components as much as possible in the content features, thus learning more general components to improve the detection generalizability.

It should be noted that we want the adversarial learning to reduce the synthesizer-related components only in the content stream, not the whole network. Therefore, when calculating the backward propagation gradients regarding $\mathcal{L}_{cls\_s}^*$, we freeze other modules and only maintain the content stream active.

\subsection{Final Classification}

We concatenate the content and synthesizer features for the final classification of the input speech. The final classification and corresponding classification loss $\mathcal{L}_{cls}$ is defined as follows:
\begin{equation}
    \begin{aligned}
        \mathbf{F}_{cls} &= \mathbf{F}_c || \mathbf{F}_s\\
        \hat{y} &= \operatorname{sigmoid}(\review{\operatorname{L_{cls}}}(\mathbf{F}_{cls}))  \\
        \mathcal{L}_{cls} &= \operatorname{BCE}(\hat{y}, y),
    \end{aligned}
    \label{eq:final_cls}
\end{equation}
where $||$ denotes the concatenation operation, \review{$\operatorname{L_{cls}}$} represents a linear classifier with weighs size of $1\times 2N$, and BCE is the binary cross-entropy loss. To highlight the discriminability of $\mathbf{F}_{cls}$ between real and fake samples, we also add a contrastive loss for $\mathbf{F}_{cls}$ as follows:
\begin{equation}
    \begin{aligned}
        \mathcal{L}_{con\_{cls}} &= CL(\{\mathbf{F}_{cls}\}_0^B, \{y\}_0^B),\\
    \end{aligned}
    \label{eq:consrastive_loss}
\end{equation}
where $B$ is the batch size.

\subsection{Synthesizer Feature Augmentation}
We propose the synthesizer feature augmentation strategy to improve the model's robustness to different synthesizer characteristics and synthesizer-content feature combinations. This strategy involves two operations: 1) randomly blends the characteristic styles within each class (real or fake) of speech features, and 2) randomly shuffles and combines the synthesizer features with the content features to simulate more feature combinations.

\subsubsection{Feature Blending}

\begin{algorithm}[!tbp]
  \caption{Feature blending.}
  \label{Alg:blending_style}
  \begin{algorithmic}[1]
      \Require Two features $\mathbf{z}_i$ and $\mathbf{z}_j$, noise level $\eta$
      \Ensure Blended feature $\mathbf{z}_i^*$
    \State $\mu_i, \sigma_i = \operatorname{mean}(\mathbf{z}_i),\operatorname{std}(\mathbf{z}_i)$
    \State $\mu_j, \sigma_j = \operatorname{mean}(\mathbf{z}_j),\operatorname{std}(\mathbf{z}_j)$
    \State $r = \operatorname{random}(0.5, 1.0)$
    \State $\mu^* = r \times \mu_i + (1-r) \times \mu_j$
    \State $\sigma^* = r \times \sigma_i + (1-r) \times \sigma_j$
    \State $\mathbf{z}_i^* = \sigma^* \times ((\mathbf{z}_i - \mu_i) / \sigma_i) + \mu^* $

    \State $r_1, r_2 = \mathcal{U}(0, \eta), \mathcal{U}(0, \eta)$ \Comment{$\mathcal{U}$ is uniform distribution.}
    \State $noise1 = r_1 * \mathcal{B}(2,5) * \mathcal{U}(-1, 1) + 1 $ \Comment{$\mathcal{B}$ is beta distribution.}
    \State $noise2 = r_2 * \mathcal{B}(2,5) * \mathcal{N}(0, 1)$ \Comment{$\mathcal{N}$ is Gaussian distribution.}
    \State $\mathbf{z}_i^* = \mathbf{z}_i^* * noise1 + noise2$
    
    \end{algorithmic}
\end{algorithm}

We randomly blend the characteristic styles within real and fake speech features separately. Given a batch of content features or synthesizer features $\mathbf{Z} =\{\mathbf{z}_1, \mathbf{z}_2, \cdots, \mathbf{z}_B\}$, we divide it into two groups: one group labeled fake and one group labeled real. For the feature $\mathbf{z}_i$ in each group, we randomly select a feature $\mathbf{z}_j$ in the same group for style blending. Algorithm~\ref{Alg:blending_style} illustrates the style blending process. We first extract the mean and standard deviation (STD) values of $\mathbf{z}_i$ and $\mathbf{z}_j$, subsequently blend these two statistics of the two samples separately, next update the feature  $\mathbf{z}_i$ using the blending mean and STD values, and finally append random noise to enhance robustness further. \review{We use Beta and Gaussian distributions to introduce noise. The Gaussian distribution effectively models natural noise and data variations, adding real-world variability to the feature spaces. The Beta distribution generates smoothing values between 0 and 1, ensuring a broad range of noise intensities.}

In the training process, we utilize the blended content feature and the blended synthesizer feature to replace the original $\mathbf{F}_c$ and $\mathbf{F}_s$ only for classification in Eq.~\eqref{eq:final_cls}. It should be noted that in Eq.~\eqref{eq:consrastive_loss}, we concatenate the unblended features to construct $\mathbf{F}_{cls}$ to compute the contrastive loss $\mathcal{L}_{con\_{cls}}$.

\subsubsection{Feature Shuffle}
In the feature shuffle process, we randomly combine the synthesizer and content features from different samples for the final fusion. Assume that $\mathbf{F}_s$ is the synthesizer feature of the sample $i$ and $\mathbf{F}_c^\prime$ is the content feature of the sample $j$ in the input batch. We combine these two features and make a prediction as follows:
\begin{equation}
    \begin{aligned}
        \hat{y}^* &= \operatorname{sigmoid}(\review{\operatorname{L_{cls}}}(\operatorname{concat}(\mathbf{F}_c^\prime, \mathbf{F}_s))).  \\
    \end{aligned}
\end{equation}
Note that these two features are blended if feature blending is used in training. For this randomly combined feature, we denote $y^*$ as its ground truth label that is real only when the samples $i$ and $j$ are both real \review{speech signals}. In a category-balanced batch, i.e., nearly equal amounts of real and fake audio, only approximately 1/4 of the new labels will be real for randomly combined feature batches. This label-imbalance issue causes the BCE loss to no longer be suitable for feature-shuffle-based classification. Therefore, we turn to focal loss (FL)~\cite{lin2017focal}, which uses two scalars to weight different categories when calculating classification loss. The FL loss is computed as follows:
\begin{equation}
    \begin{aligned}
        \mathcal{L}_{cls\_aug} &= \operatorname{FL}(\hat{y}^*, y^*)\\
        & = -\alpha\left(1-\hat{y}^*\right)^\gamma \log \left(\hat{y}^*\right)
    \end{aligned}
\end{equation}
where $\alpha$ and $\gamma$ are set to 0.25 and 2 by default, respectively.

\subsection{Loss Function}

Our final loss function comprises the main classification loss and all the regularization losses. We define the total loss for model training as follows:
\begin{equation}
    \begin{aligned}
        \mathcal{L} = &( \mathcal{L}_{cls} + \beta_0 * \mathcal{L}_{cls\_aug} \\
        & + \beta_1 * (\mathcal{L}_{cls\_s} + 0.5 * \mathcal{L}_{con\_s}) \\
        & + \beta_2 * (\mathcal{L}_{cls\_c} + \mathcal{L}_{cls\_s}^*) + 
        \beta_3 * \mathcal{L}_{con\_{cls}})
    \end{aligned}
    \label{eq:loss_function}
\end{equation}
where $(\beta_0, \beta_1, \beta_2, \beta_3)$ are adjustment parameters for sub-losses.

\section{Experiments}
\label{sec:experiments}
\subsection{Datasets}

\begin{table}[!tbp]
    \centering
    \caption{Details of three audio deepfake datasets: WaveFake, LibriSeVoc, and DECRO. EN, JP, and ZH donate the English, Japanese, and Chinese subsets, respectively.}
    \label{tab:datasets}
    \small
    \setlength\tabcolsep{3pt}
    \begin{tabular}{cccccccc}
        \toprule
        \multirow{2}{*}{Details}     & \multicolumn{2}{c}{WaveFake} & LibriSeVoc & \multicolumn{2}{c}{DECRO}\\ \cmidrule(lr){2-3} \cmidrule(lr){4-4} \cmidrule(l){5-6}
        & EN & JP & EN & EN & ZH \\
        \midrule
        \tabincell{c}{No.\\ Synthesizers} & 8 & 2 & 6 & 10 & 10 \\
        No. Real & 13100 & 5000 & 13201 & 12484 & 21218 \\
        No. Fake  & \review{13100$\times$8} & \review{5000$\times$2} & \review{13201$\times$6} & 42799 & 41880 \\
        \bottomrule
             
    \end{tabular}
\end{table}

\begin{table*}[!tbp]
    \centering
    \small
    \caption{Inner evaluation on the LibriSeVoc, WaveFake, and DECRO datasets. We report the AUC ($\uparrow$) / EER ($\downarrow$) ($\%$) performance as the evaluation metrics for each method. The best scores are formatted to \best{red}, and the second best scores are formatted to \secondbest{violet}.}
    \label{tab:inner-evaluation}
    \color{TableColor}
    \begin{tabular}{lcrrrrrrrrrrrrrrrrrr}
    \toprule
    \multirow{2.5}{*}{Method} & \multirow{2.5}{*}{Input} & \multicolumn{4}{c}{Inner-Dataset Evaluation} & \multirow{2.5}{*}{\textbf{\textit{Average}}} \\ \cmidrule(lr){3-6}  
    &       &    \multicolumn{1}{c}{LibriSeVoc} &  \multicolumn{1}{c}{WaveFake}  &  \multicolumn{1}{c}{DECRO-EN} &  \multicolumn{1}{c}{DECRO-ZH} &  \\
    \midrule

LCNN & LFCC &   99.96/0.90 & 99.98/0.64 & 99.96/0.90 & 99.88/1.43 & 99.94/0.97 \\
RawNet2 & Raw & 95.00/6.94 & 97.93/6.95 & 99.37/3.68 & 99.32/3.85 & 97.91/5.36 \\
RawGAT &Raw &  99.89/1.45 & 99.92/1.25 & 99.89/1.45 & 99.87/1.56 & 99.89/1.43 \\
Wave2Vec2 &Raw &  \best{100.00}/\secondbest{0.09} & 99.99/0.44 & \secondbest{99.99}/\secondbest{0.47} & 99.98/\secondbest{0.43} & \best{99.99}/0.35 \\
WaveLM & Raw &   \best{100.00}/\best{0.03} & \best{100.00}/\secondbest{0.26} & 99.98/0.55 & \best{99.99}/\best{0.41} & \best{99.99}/\secondbest{0.31} \\
RawNet2-Voc & Raw & 99.44/2.86 & 99.37/3.93 & 99.42/3.42 & 99.10/4.36 & 99.33/3.64 \\
AudioClip & Raw &  99.32/3.98 & 99.92/1.29 & 99.88/0.91 & 99.58/2.85 & 99.67/2.26 \\
Wav2Clip &Log-Scale Spec & 99.83/1.60 & 99.99/0.30 & 99.98/0.68 & 99.21/4.08 & 99.75/1.66 \\
AASIST & Raw &  99.91/1.33 & 99.84/1.60 & 99.91/1.35 & 99.56/2.92 & 99.81/1.80 \\
SFATNet & Log-Scale Spec & 98.33/5.95 & 99.89/1.52 & 99.85/1.75 & 97.50/8.17 & 98.89/4.35 \\
ASDG &  LFCC &   99.93/1.22 & 99.94/0.77 & 99.92/1.10 & 99.80/1.76 & 99.90/1.21 \\
\rowcolor{LightCyan}
\textbf{Ours} & Log-Scale Spec &  99.99/0.43 & \best{100.00}/\best{0.07} & \best{100.00}/\best{0.21} & \best{99.99}/0.45 & \best{99.99}/\best{0.30} \\
    \bottomrule
    \end{tabular}
\end{table*}

We evaluate our proposed method using three audio deepfake datasets:  WaveFake~\cite{frank2021wavefake}, LibriSeVoc~\cite{sunAISynthesizedVoiceDetection2023-LibraSeVoc} and DECRO~\cite{ba2023transferring-DECRO}. Table~\ref{tab:datasets} lists the number of used synthesizer methods, languages, and real and fake \review{speech signals} of these datasets.

\textbf{WaveFake.} This dataset was generated utilizing \review{a TTS synthesizer} and seven pre-trained synthesizer methods: Multi-band MelGAN (MB-MelGAN), Parallel WaveGAN (PWG)~\cite{yamamotoParallelWaveganFast2020}, Full-band MelGAN (FB-MelGAN), HIFI-GAN~\cite{kong2020hifi}, MelGAN~\cite{kumar2019melgan}, MelGAN-large (MelGAN-L), WaveGlow. The authors employed these \review{seven non-TTS} synthesizers on two reference datasets, English (EN) and Japanese (JP), to generate synthetic \review{speech signals}. Specifically, the authors built the EN subset based on the LJSpeech~\cite{ljspeech17} corpus using these seven synthesizers and built the JP subset based on the basic5000 corpus of JUST~\cite{JSUT} dataset using only the MB-MelGAN and PWG. It should be noted that the above synthetic \review{speech signals} were produced by feeding the mel-spectrograms of raw waveforms into different vocoders, i.e., they are self-vocoding samples. To synthesize samples from a complete TTS pipeline, the authors employed a Conformer network~\cite{gulatiConformerConvolutionaugmentedTransformer2020} to map non-LJSpeech \review{speech signals} to mel-spectrograms, which were fed into a fine-tuned PWG model to produce synthetic \review{speech signals}. We denote this subset as TTS in the following sections.

\textbf{LibriSeVoc.} The authors employed six synthesizers, WaveNet, WaveRNN, WaveGrad, MelGAN, PWG, and  DiffWave, on the LibriTTS~\cite{zen2019libritts} corpus to build this dataset. Concretely, the authors randomly selected 13,201 EN audio clips from LibriTTS as references and employed each synthesizer to generate corresponding synthetic \review{speech signals}. Similar to the WaveFake dataset, the synthetic \review{speech signals} in this dataset are also self-vocoding samples.

\textbf{DECRO.} It contains EN and Chinese (ZH) subsets. The authors built each subset using the same ten types of synthetic algorithms: HIFI-GAN, MB-MelGAN, PWG, Tacotron, FastSpeech2, StarGANv2, VITS~\cite{kim2021conditional-VITS}, NVCNet, Baidu, Xunfei. In contrast to the WaveFake and LibriSeVoc datasets, the synthetic audios in the DECRO were all generated by TTS or VC.

When splitting these datasets for training/validation/testing, we apply custom splits on the WaveFake and LibriSeVoc datasets since they do not have standard splits. For the DECRO dataset,  we follow its publicly available standard splits\footnote{https://github.com/petrichorwq/DECRO-dataset\#division}.

\subsection{Comparison Methods}

In addition to the AASIST~\cite{jung2022aasist}, RawNet2-Voc~\cite{sunAISynthesizedVoiceDetection2023-LibraSeVoc}, \review{SFATNet~\cite{cuccovillo_audio_2023_f0} and ASDG~\cite{xie2024domain_ASDG}} introduced in Section~\ref{sec:related_work_deepfake}, we also compare our method with the following methods:
\begin{itemize}

    \item LFCC-LCNN~\cite{lavrentyeva2019stc-LCNN}: It is a classical model for speech-related tasks and is one of the baselines in the ASVspoof 2021 challenge. It builds a light CNN architecture and utilizes the LFCC features as the input for speech detection. For simplicity, we refer to it as LCNN hereafter.

    \item RawNet2~\cite{jung2020improved-RawNet2}: It is also one of the baselines in the ASVspoof 2021 challenge. To address the limitations of traditional feature-based approaches, this model constructs a 1D CNN architecture and learns features directly from raw waveforms.

    \item RawGAT~\cite{tak21_asvspoof_RawGAT}: It is a spectro-temporal graph attention network that learns the relationship between cues spanning different sub-bands and temporal intervals from raw waveforms.

    \item Wav2Vec2~\cite{baevski2020wav2vec}, WavLM~\cite{chen2022wavlm}. They are large-scale pre-training models that learn universal speech signal representations from large-scale unlabeled \review{speech signals} and can be adapted to full-stack downstream speech tasks.

    \item Wav2Clip~\cite{wu2022wav2clip}. Distilling from the Contrastive Language-Image Pre-training (CLIP) model, this model projects audio into a shared embedding space with images and text to learn robust audio representation and can be fine-tuned to downstream audio tasks.
    
    \item AudioClip~\cite{guzhov2022audioclip}: It is an extension of the CLIP model that incorporates the ESResNeXt~\cite{guzhovESResNeTfbspLearning2021} audio model into the CLIP framework to learn audio representation. 
    
\end{itemize}

We use the publicly available codes of these methods for training and testing. For pre-training methods Wav2Vec2, WavLM, Wav2Clip, and AudioClip, we utilize their publicly available pre-trained models to initialize the model weights and fine-tune them for speech deepfake detection.

\subsection{Data Preprocessing}
We convert all audio clips to mono-channel with a sampling rate of 16kHz. We set the length of the audio clip as 48000 (three seconds) in training, validation, and testing. The audio clip is padded with itself if its length is shorter than three seconds. For those audio clips longer than three seconds, we randomly crop a three-second segment at each reading time in the training process and crop only the middle three-second segment for validation and testing. Considering that these three datasets all have more fake samples than real samples, we employ over-sampling to the real category to solve the imbalance issue in the training process.

\begin{table*}[t]
    \centering
    \small
    \caption{Cross-method evaluation on LibriSeVoc dataset. We train all the models on the deepfake \review{speech signals} generated by MelGAN and PWG methods and test their  AUC($\uparrow$) / EER ($\downarrow$) ($\%$) performance on other synthesizer methods.}
    \label{tab:cross-method-LibriSeVoc}
    \color{TableColor}
    \begin{tabular}{lrrrrrrrrrrrrrrrrrrr}
    \toprule
    Method &  \multicolumn{1}{c}{DiffWave} &  \multicolumn{1}{c}{WaveNet} &  \multicolumn{1}{c}{WaveRNN} &  \multicolumn{1}{c}{\tabincell{c}{WaveGrad}} &  \multicolumn{1}{c}{\textbf{\textit{Average}}} \\
    \midrule
LCNN & 97.52/\phantom{0}7.93 & 69.18/36.43 & 74.52/31.76 & 99.33/\phantom{0}3.64 & 85.14/19.94 \\
RawNet2 & 74.63/31.35 & 78.26/28.13 & 65.79/38.44 & 78.56/28.23 & 74.31/31.54 \\
RawGAT & 75.82/30.78 & 77.19/29.22 & 74.25/32.18 & 85.69/21.53 & 78.24/28.43 \\
Wave2Vec2 & \secondbest{98.17}/\secondbest{\phantom{0}4.82} & \best{99.98}/\best{\phantom{0}0.42} & 94.40/11.09 & 97.33/\phantom{0}6.24 & \secondbest{97.47}/\best{\phantom{0}5.64} \\
WaveLM & 96.43/\phantom{0}6.25 & \secondbest{99.51}/\secondbest{\phantom{0}1.39} & \secondbest{94.80}/\secondbest{11.04} & 96.28/\phantom{0}6.21 & 96.76/\phantom{0}6.22 \\
RawNet2-Voc & 67.29/36.58 & 68.89/34.94 & 62.76/39.82 & 69.64/34.79 & 67.15/36.53 \\
AudioClip & 89.87/18.36 & 87.91/19.86 & 71.09/34.16 & 93.81/13.89 & 85.67/21.57 \\
Wav2Clip & 94.30/13.02 & 79.51/28.00 & 84.10/24.13 & 75.02/31.45 & 83.23/24.15 \\
AASIST & 77.12/29.68 & 74.42/31.67 & 77.29/29.67 & 89.67/18.61 & 79.62/27.41 \\
SFATNet & 92.34/15.03 & 86.56/20.65 & 79.31/28.05 & 96.55/\phantom{0}9.34 & 88.69/18.27 \\
ASDG & 98.84/\phantom{0}6.43 & 81.28/22.58 & 84.61/21.88 & \best{99.84}/\best{\phantom{0}1.59} & 91.14/13.12 \\
\textbf{Ours} & \best{99.11}/\best{\phantom{0}3.99} & 97.07/\phantom{0}8.10 & \best{95.19}/\best{10.44} & \secondbest{99.79}/\secondbest{\phantom{0}1.98} & \best{97.79}/\secondbest{\phantom{0}6.12}  \\
    \bottomrule
    \end{tabular}
\end{table*}

\begin{figure*}
    \centering
    \small

    \begin{minipage}[b]{0.95\linewidth}
        \centerline{\includegraphics[width=1\linewidth]{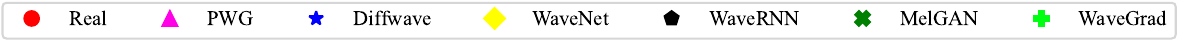}}
    \end{minipage}

    \begin{minipage}[b]{0.99\linewidth}
    \centering
        \begin{minipage}[b]{0.16\linewidth}
            \centerline{\includegraphics[width=1\linewidth]{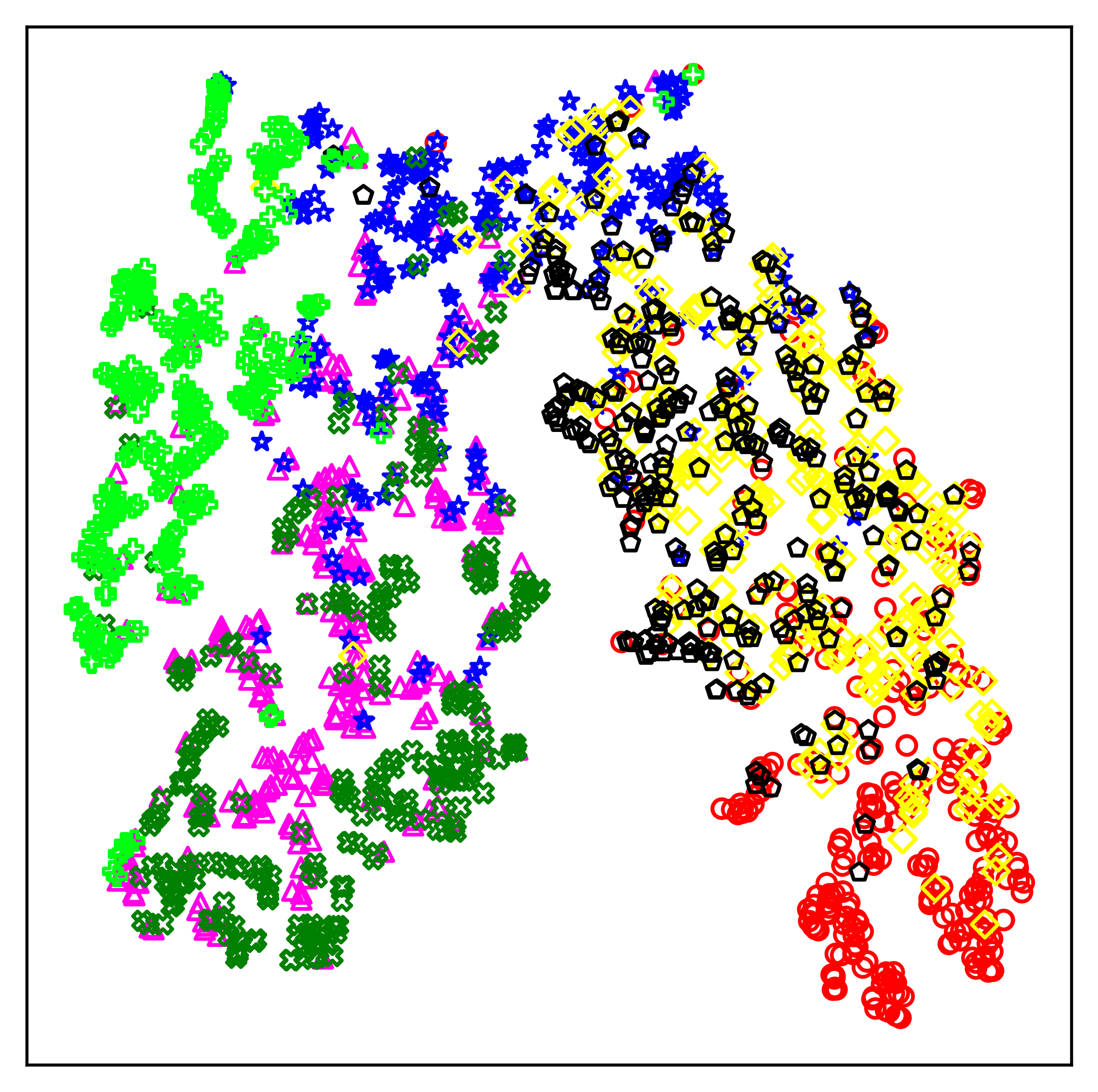}}
            \centerline{LCNN~\cite{lavrentyeva2019stc-LCNN}}
        \end{minipage}
        \begin{minipage}[b]{0.16\linewidth}
            \centerline{\includegraphics[width=1\linewidth]{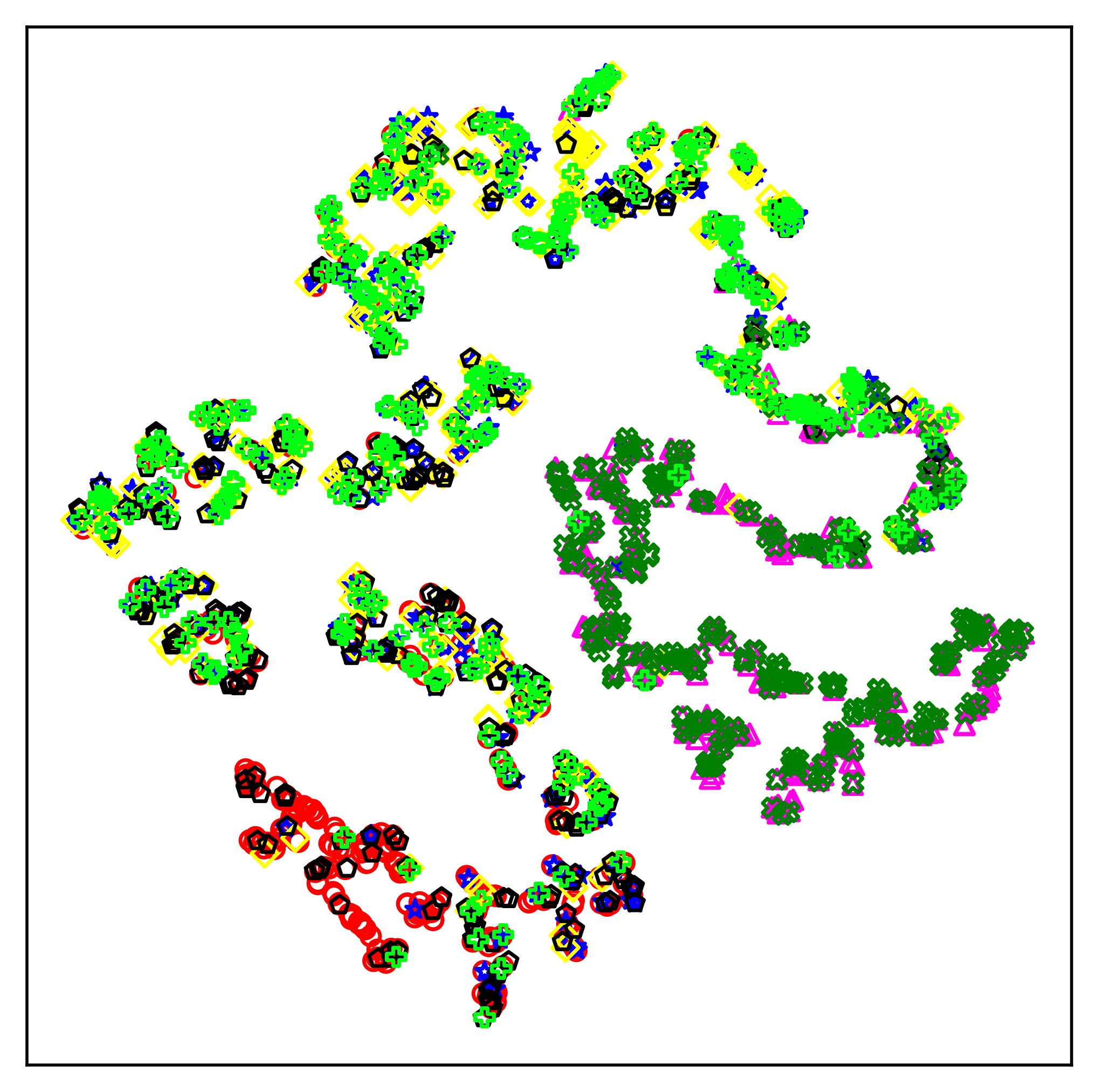}}
            \centerline{RawNet2~\cite{jung2020improved-RawNet2}}
        \end{minipage}
        \begin{minipage}[b]{0.16\linewidth}
            \centerline{\includegraphics[width=1\linewidth]{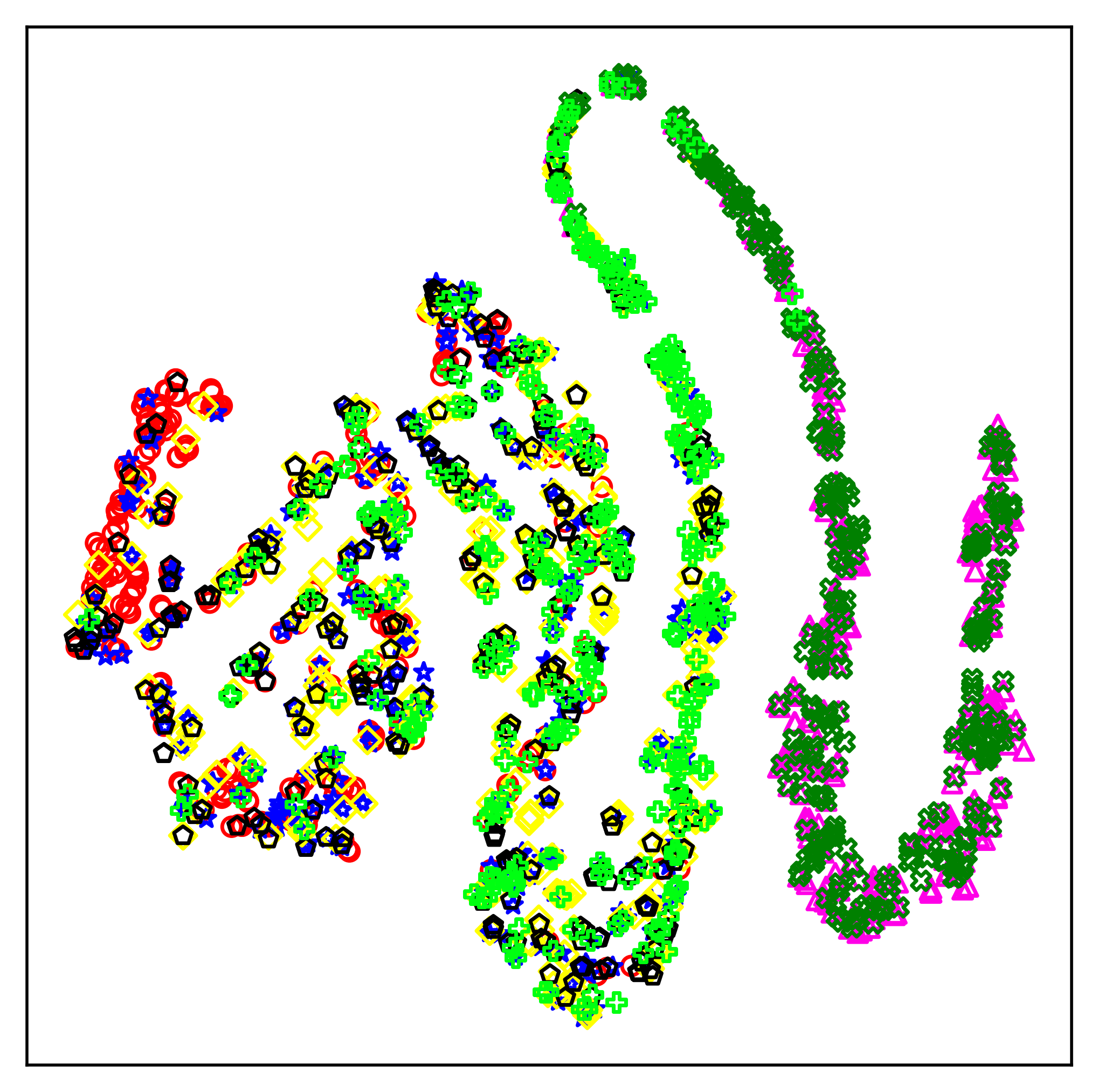}}
            \centerline{RawGAT~\cite{tak21_asvspoof_RawGAT}}
        \end{minipage}
        \begin{minipage}[b]{0.16\linewidth}
            \centerline{\includegraphics[width=1\linewidth]{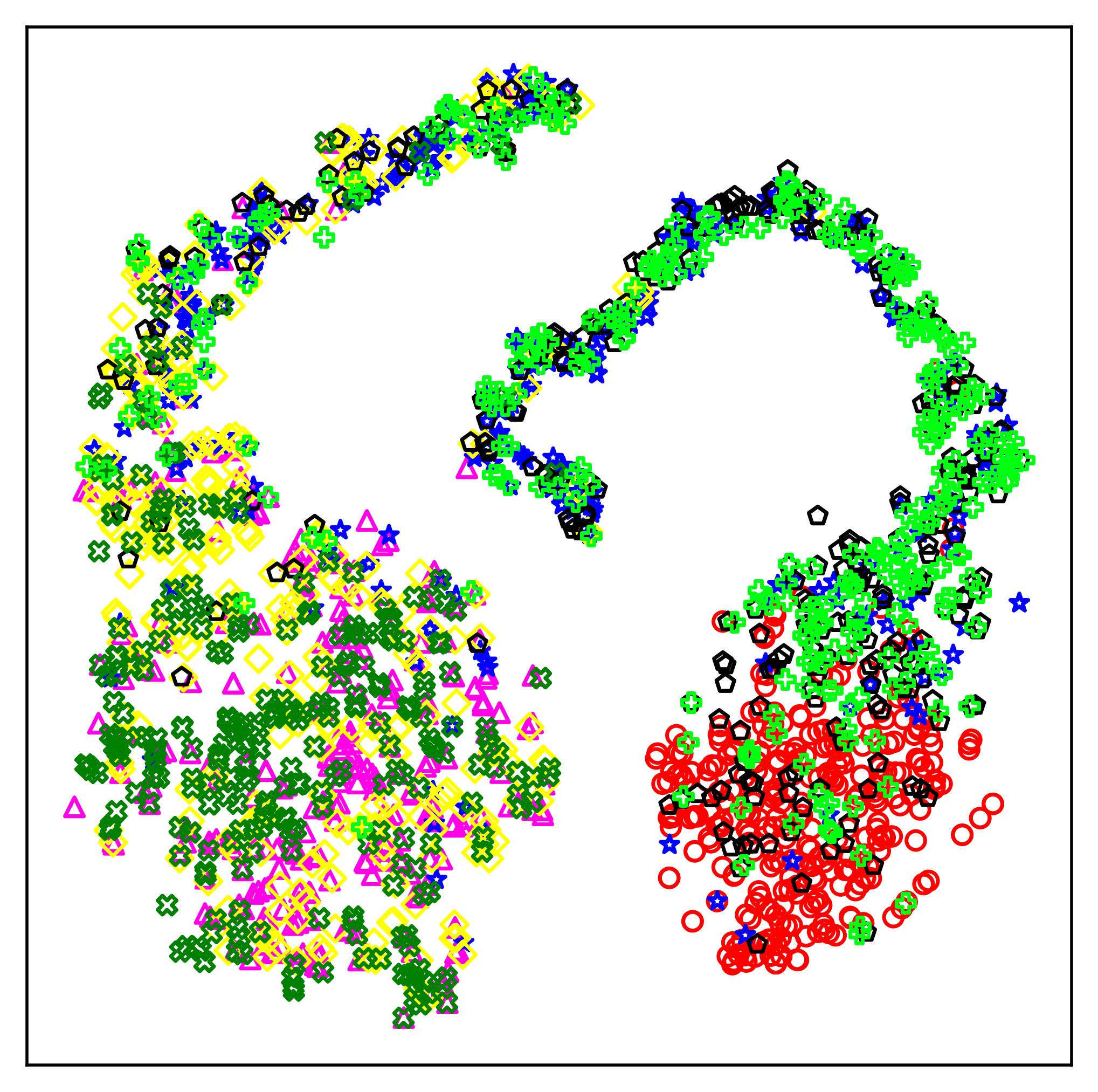}}
            \centerline{Wave2Vec2~\cite{baevski2020wav2vec}}
        \end{minipage}
        \begin{minipage}[b]{0.16\linewidth}
            \centerline{\includegraphics[width=1\linewidth]{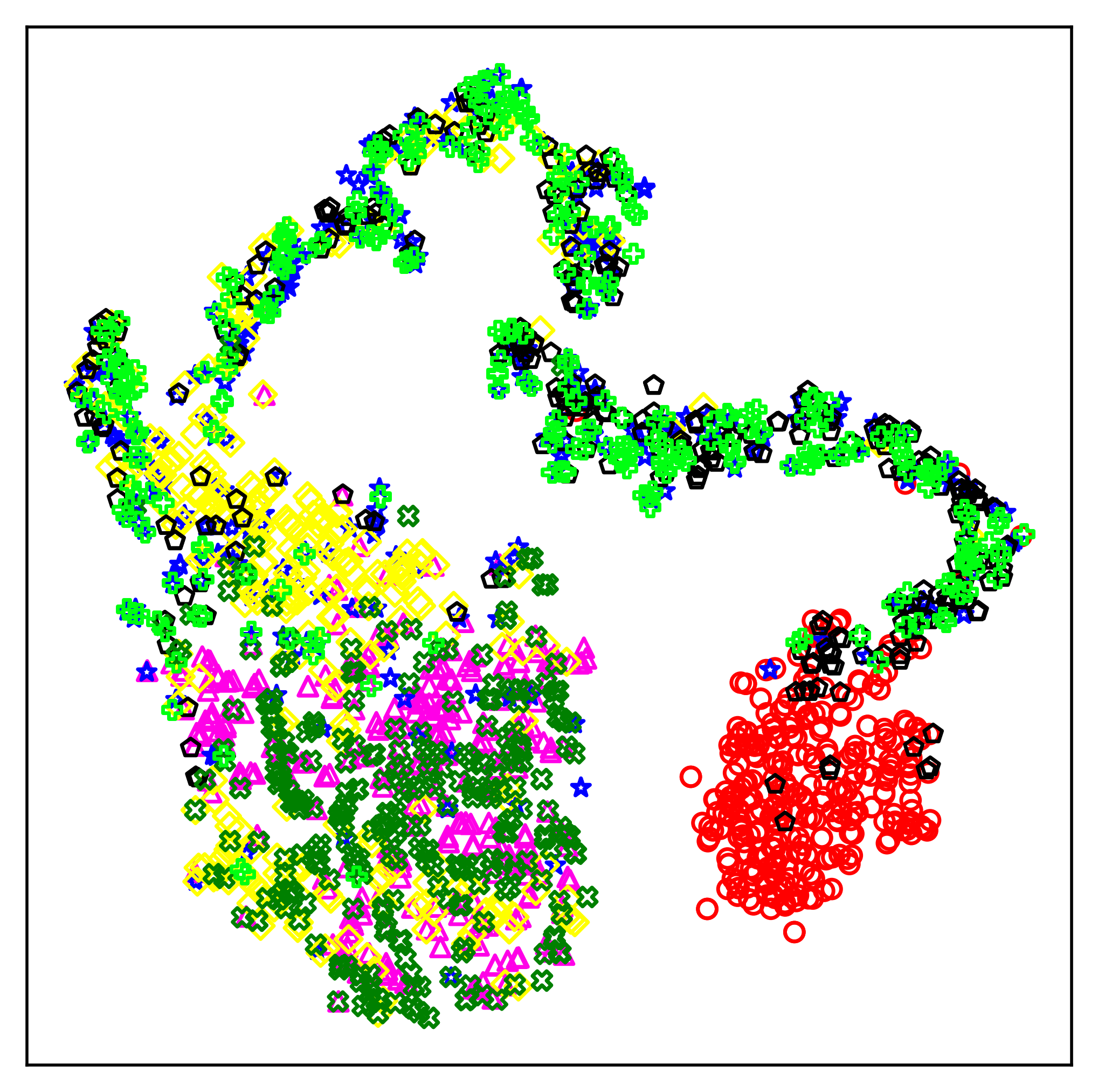}}
            \centerline{WaveLM~\cite{chen2022wavlm}}
        \end{minipage}
        \begin{minipage}[b]{0.16\linewidth}
            \centerline{\includegraphics[width=1\linewidth]{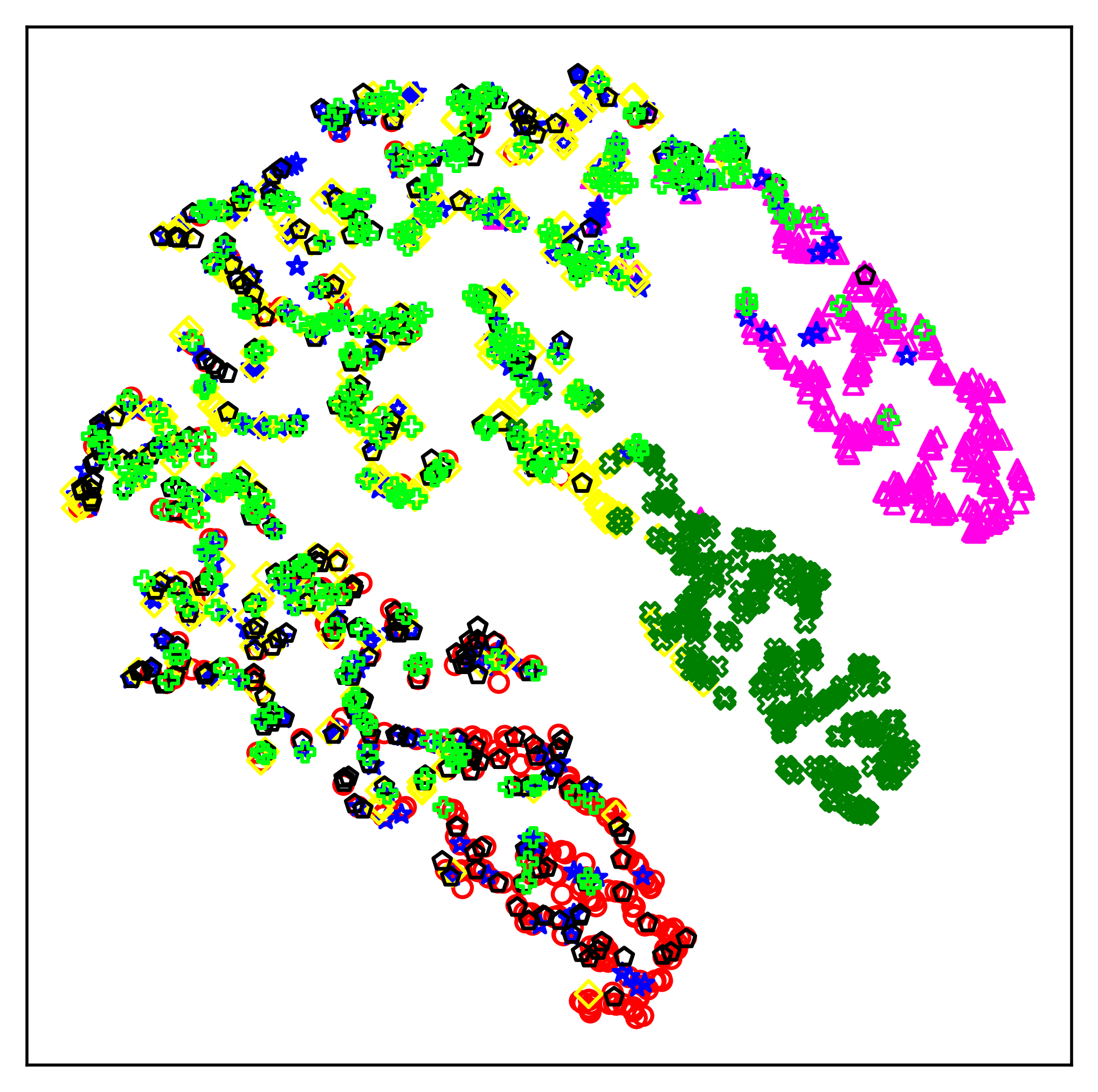}}
            \centerline{RawNet2-Voc~\cite{sunAISynthesizedVoiceDetection2023-LibraSeVoc}}
        \end{minipage}

        \vspace{5pt}
        
        \begin{minipage}[b]{0.16\linewidth}
            \centerline{\includegraphics[width=1\linewidth]{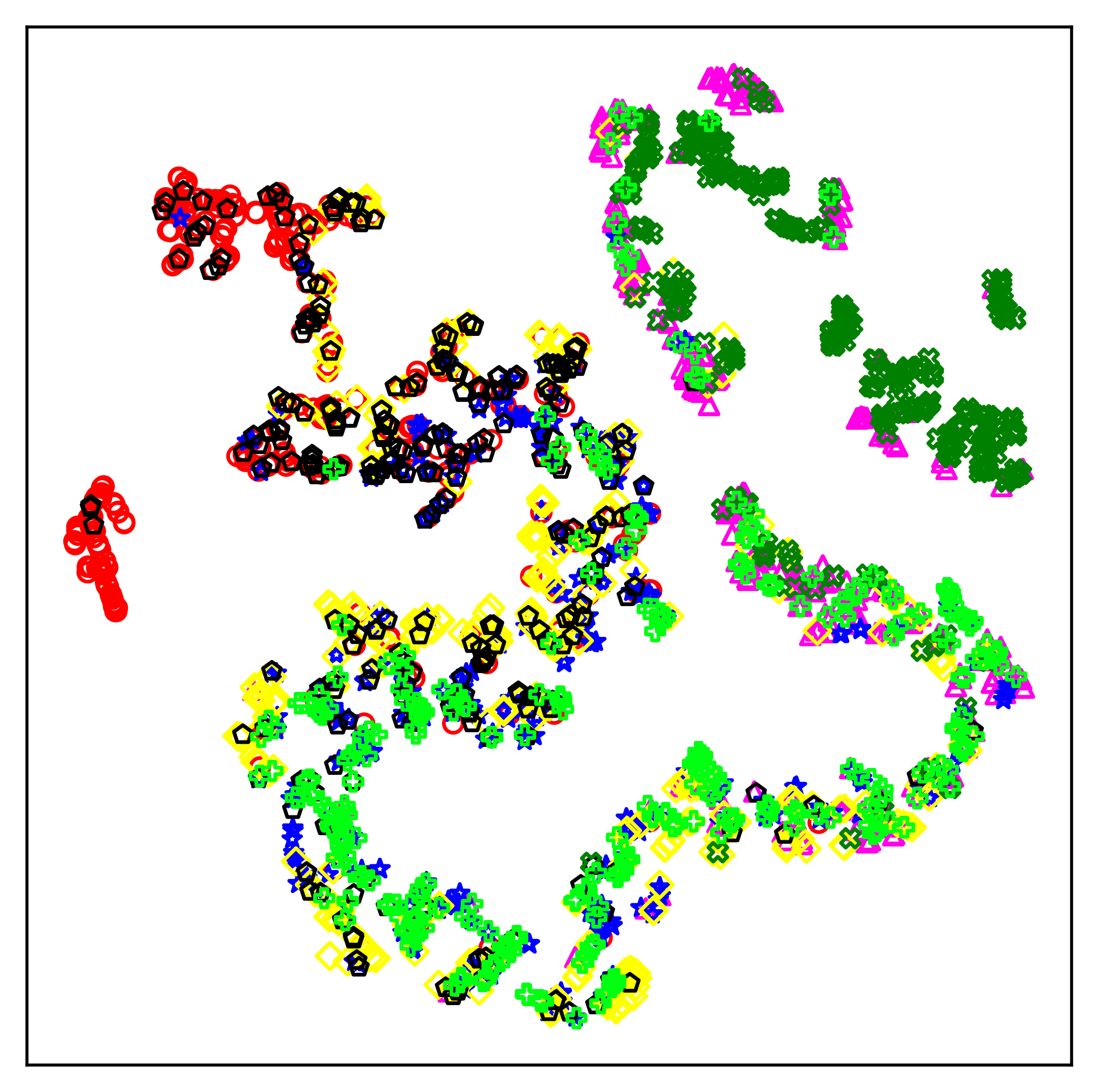}}
            \centerline{AudioClip~\cite{guzhov2022audioclip}}
        \end{minipage}
        \begin{minipage}[b]{0.16\linewidth}
            \centerline{\includegraphics[width=1\linewidth]{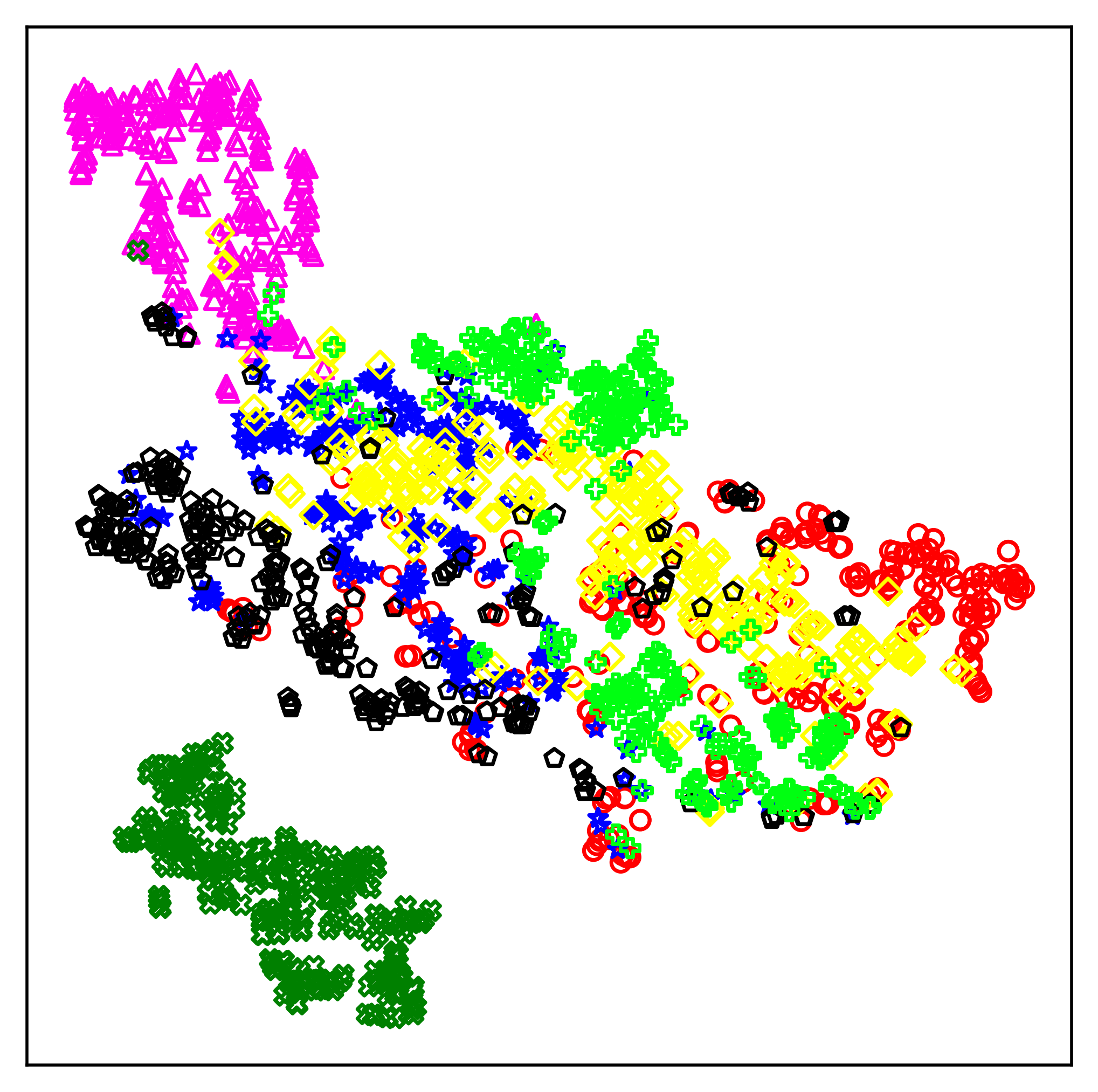}}
            \centerline{Wav2Clip~\cite{wu2022wav2clip}}
        \end{minipage}
        \begin{minipage}[b]{0.16\linewidth}
            \centerline{\includegraphics[width=1\linewidth]{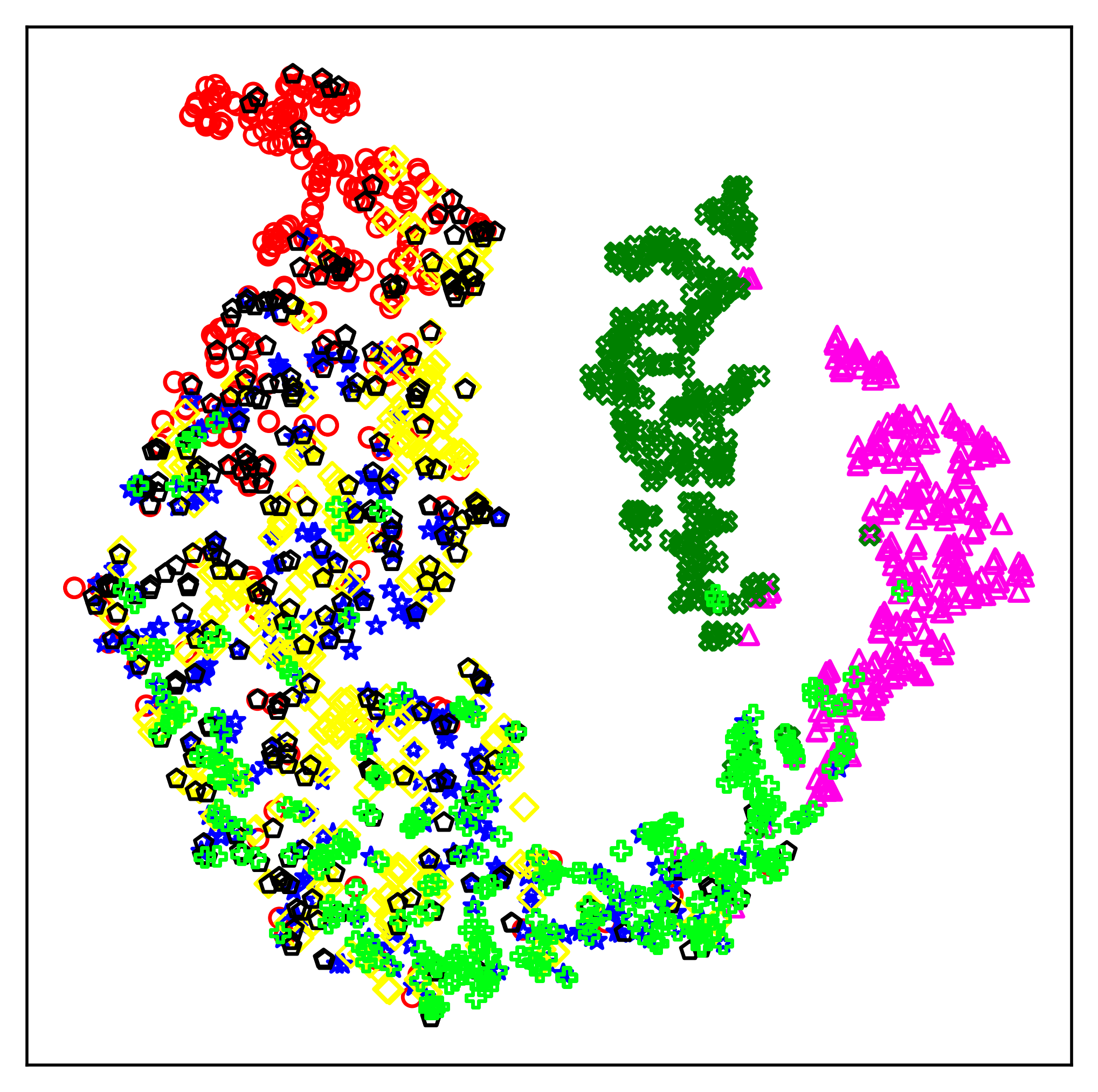}}
            \centerline{AASIST~\cite{jung2022aasist}}
        \end{minipage}
        \begin{minipage}[b]{0.16\linewidth}
            \centerline{\includegraphics[width=1\linewidth]{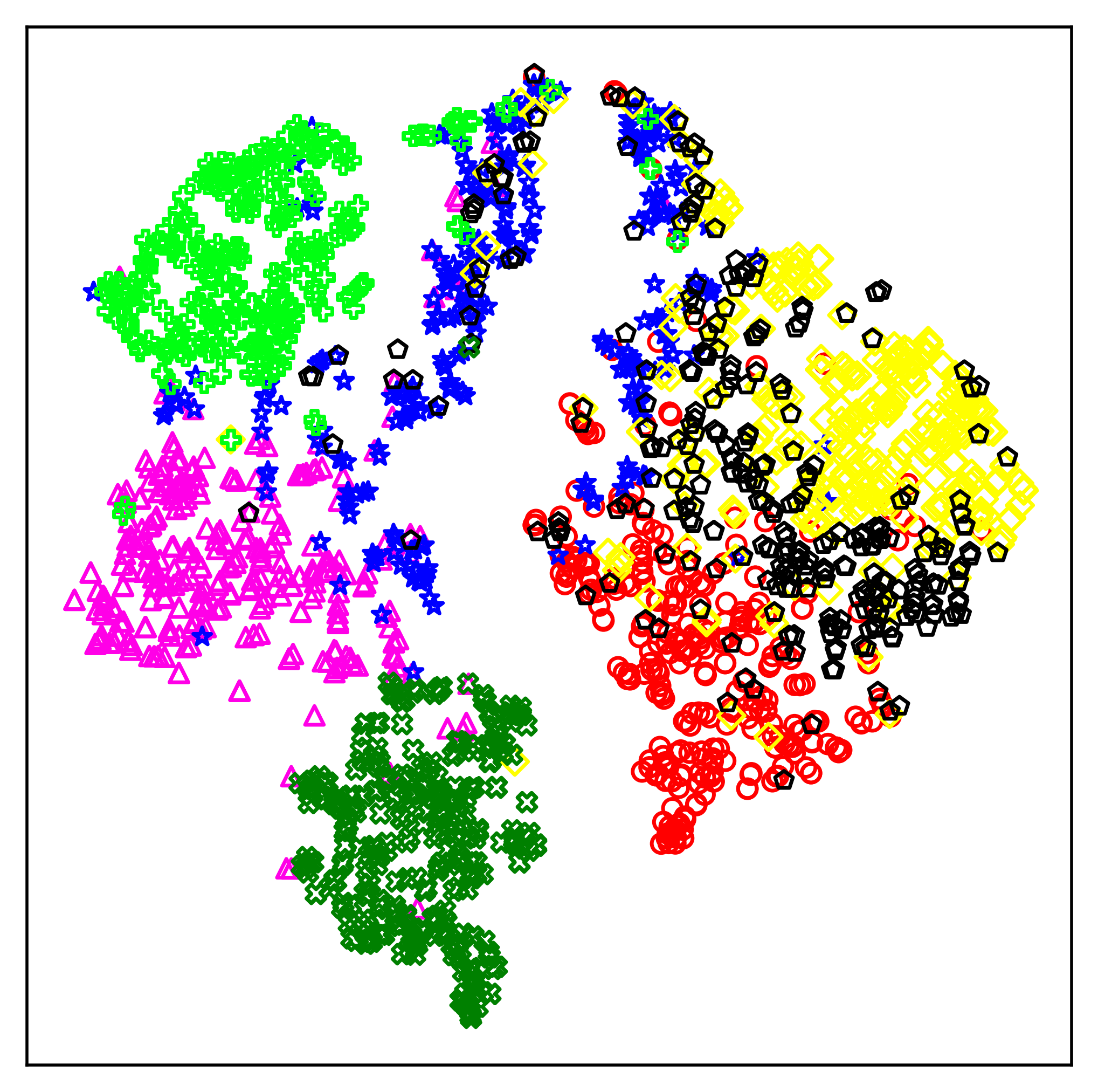}}
            \centerline{\review{ASDG}~\cite{xie2024domain_ASDG}}
        \end{minipage}
        \begin{minipage}[b]{0.16\linewidth}
            \centerline{\includegraphics[width=1\linewidth]{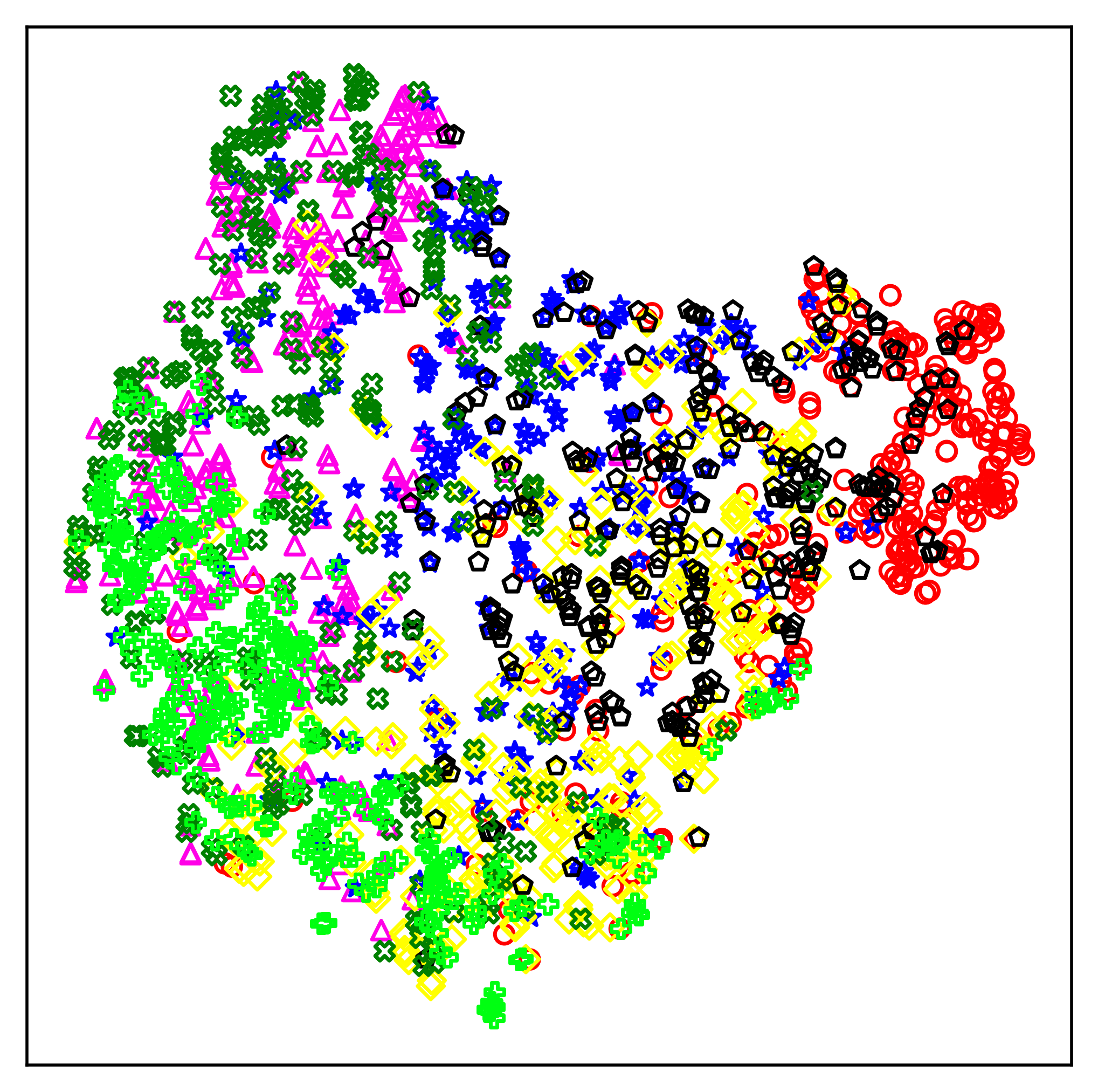}}
            \centerline{\review{SFATNet}~\cite{cuccovillo_audio_2023_f0}}
        \end{minipage}
        \begin{minipage}[b]{0.16\linewidth}
            \centerline{\includegraphics[width=1\linewidth]{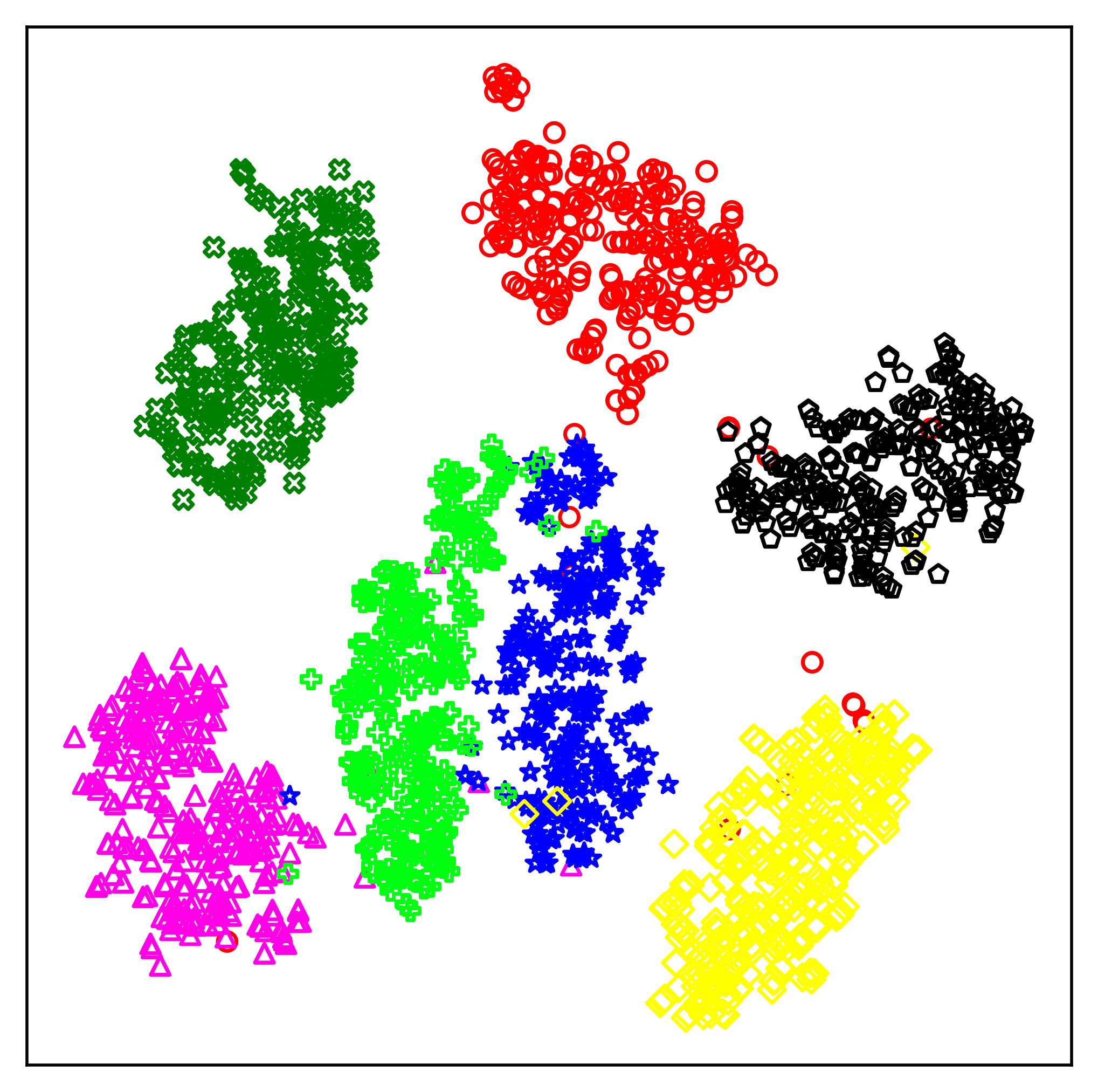}}
            \centerline{\textbf{Ours}}
        \end{minipage}
    \end{minipage}

    \caption{T-SNE visualization in the cross-evaluation task on the LibriseVoc dataset. For each deepfake speech detection method, we extract the latent features from the validation and test subsets and randomly extract 300 samples of the real and each fake method for visualization.}
    \label{fig:t_SNE_cross_method}
\end{figure*}

\subsection{Implementation Details}
We set the numbers of feature channels to (64, 128, 256, 512) in the four convolutional blocks of ResNet18. We initialize the weights of our network using the pre-training ResNet18 in Wav2Clip~\cite{wu2022wav2clip}. In our pseudo-labeling-based supervised learning, the speech transformation has $N_1 = 10$ compression settings and $N_2=16$ speed settings. Concretely, the compression transform involves three codecs (aac, ops, mp3) and three bitrates (16000, 32000, 64000), while the speed transform is in the range of 0.5 to 2.0. We set the noise level \review{$\eta=10$} in our feature blending strategy.

The $\alpha$ in the contrastive loss is set to 0.4. The $(\beta_0, \beta_1, \beta_2, \beta_3)$ in the loss function are set to (1.0, 0.5, 0.5, 0.5) by default. We set the batch size as 128 and utilize the Adam optimizer~\cite{yong2020gradient} to optimize the model parameters. The learning rate is set to 0.0001 with a weight decay of 0.01. 
We use the PyTorch framework to implement all the methods and conduct all the experiments on a GTX 4090 GPU device. The early stopping strategy is used to terminate model training when the area under the ROC Curve (AUC) performance no longer improves within three epochs.

\section{Experiments Results}
To demonstrate the robust detection performance of our method, we evaluate our method and the comparison methods on the inner-dataset, cross-method, cross-dataset, and cross-language scenarios. We measure the model performance using the AUC and equal error rate (EER)~\cite{ASVspoof2021} metrics.  \review{Note that for each detection method, we train it with ten runs in each task, where each run utilizes a different global random seed to control model initialization and data loading during training. Then, we report the average values on the ten runs.}

\subsection{Inner-Dataset Evaluation}

\begin{table*}[t]
    \centering
    \footnotesize
    \caption{Cross-method evaluation on WaveFake dataset. We train all the models on the deepfake speech samples generated by MelGAN and PWG methods and test their  AUC($\uparrow$) / EER($\downarrow$) ($\%$) performance on other synthesizer methods.}
    \label{tab:cross-method-WaveFake}
    \color{TableColor}
    \begin{tabular}{lrrrrrrrrrrrrr}
    \toprule
    \multirow{2}{*}{Method} & \multicolumn{5}{c}{Self-Vocodered} & \multirow{2}{*}{TTS} & \multirow{2}{*}{\textbf{\textit{Average}}} \\ \cmidrule(lr){2-6}
     &  \multicolumn{1}{c}{MB-MelGAN} &  \multicolumn{1}{c}{FB-MelGAN} &  \multicolumn{1}{c}{HIFIGAN} &  \multicolumn{1}{c}{MelGAN-L} &  \multicolumn{1}{c}{WaveGlow} &  &  \\
    \midrule
LCNN & 99.67/\phantom{0}2.60 & 99.75/\phantom{0}2.18 & 98.88/\phantom{0}5.02 & \best{100.00}/\secondbest{\phantom{0}0.07} & 99.08/\phantom{0}4.51 & 99.99/\phantom{0}0.46 & 99.56/\phantom{0}2.47 \\
RawNet2 & 70.29/35.28 & 65.53/39.05 & 63.14/40.78 & 99.68/\phantom{0}2.67 & 86.63/21.44 & 84.62/23.36 & 78.31/27.10 \\
RawGAT & 96.27/\phantom{0}7.48 & 92.94/13.84 & 92.43/14.47 & 96.63/\phantom{0}4.42 & 95.88/\phantom{0}7.06 & 98.33/\phantom{0}5.41 & 95.41/\phantom{0}8.78 \\
Wave2Vec2 & 95.58/\phantom{0}5.58 & 93.91/\phantom{0}9.43 & 92.47/11.77 & 95.70/\phantom{0}4.94 & 96.58/\phantom{0}4.76 & 95.95/\phantom{0}6.12 & 95.03/\phantom{0}7.10 \\
WaveLM & \secondbest{99.99}/\secondbest{\phantom{0}0.30} & 99.84/\phantom{0}1.75 & 99.12/\phantom{0}4.21 & \best{100.00}/\phantom{0}0.22 & \secondbest{99.98}/\secondbest{\phantom{0}0.51} & 99.89/\phantom{0}1.35 & 99.80/\secondbest{\phantom{0}1.39} \\
RawNet2-Voc & 63.32/39.90 & 61.47/41.24 & 59.22/43.05 & 98.83/\phantom{0}5.01 & 79.14/27.61 & 73.84/32.11 & 72.64/31.49 \\
AudioClip & 99.87/\phantom{0}1.61 & 99.64/\phantom{0}2.89 & 98.91/\phantom{0}5.25 & 99.99/\phantom{0}0.26 & 96.72/\phantom{0}9.67 & 99.86/\phantom{0}1.58 & 99.16/\phantom{0}3.55 \\
Wav2Clip & 99.79/\phantom{0}2.02 & 99.88/\phantom{0}1.41 & 98.96/\phantom{0}4.81 & 99.99/\phantom{0}0.25 & 99.58/\phantom{0}3.01 & \secondbest{99.96}/\phantom{0}0.76 & 99.69/\phantom{0}2.04 \\
AASIST & 98.99/\phantom{0}4.86 & 94.30/13.04 & 93.14/14.61 & 99.99/\phantom{0}0.25 & 99.25/\phantom{0}3.86 & 99.67/\phantom{0}2.25 & 97.56/\phantom{0}6.48 \\
SFATNet & 72.21/33.60 & 70.61/34.88 & 70.34/35.12 & 71.71/33.89 & 70.85/34.61 & 67.16/37.29 & 70.48/34.90 \\
ASDG & 99.83/\phantom{0}2.25 & \secondbest{99.90}/\secondbest{\phantom{0}1.11} & \secondbest{99.71}/\secondbest{\phantom{0}3.64} & 99.95/\phantom{0}0.16 & 99.79/\phantom{0}2.88 & 99.95/\secondbest{\phantom{0}0.18} & \secondbest{99.86}/\phantom{0}1.70 \\
\textbf{Ours} & \best{100.00}/\best{\phantom{0}0.19} & \best{99.98}/\best{\phantom{0}0.48} & \best{99.70}/\best{\phantom{0}2.32} & \best{100.00}/\best{\phantom{0}0.01} & \best{100.00}/\best{\phantom{0}0.11} & \best{100.00}/\best{\phantom{0}0.00} & \best{99.95}/\best{\phantom{0}0.52} \\
    \bottomrule
    \end{tabular}
\end{table*}

The training/validation/testing subsets in inner-dataset evaluation tasks consist of the same synthesizer methods for all the datasets. Specifically, we split the training/validation/testing subsets at the rate of 0.6/0.2/0.2 for the WaveFake and LibriseVoc datasets. \review{Each genuine file in the datasets has a unique ID. We first split the file IDs and then assign the real samples, along with their corresponding deepfake samples, to each subset based on these IDs. This ensures that each subset maintains class balance across the various synthesizer methods, with consistent synthesizer methods across the subsets.}

Table~\ref{tab:inner-evaluation} lists the inner evaluation results on the LibriSeVoc, WaveFake, and DECRO datasets. As can be seen, all comparison methods and our method demonstrate high-level detection performance. Our method \review{achieves the best average AUC and EER scores, with values of $99.99\%$ and $0.30\%$, respectively.}

All the deepfake speech detection methods can achieve more than $97\%$ on the AUC scores and less than $6\%$ on the EERs. This high-level performance is because all these methods have strong learning capabilities and can achieve good detection performance when the training and test data have the same distribution. Therefore, detection generalizability is a more important metric for deepfake speech detection methods, i.e., better performance even on unseen data distributions.

\subsection{Cross-Method Evaluation}

\subsubsection{LibriSeVoc and WaveFake}

We evaluate the cross-method ability of all deepfake speech detection methods on the LibriSeVoc and WaveFake datasets. For each dataset, we train and validate detection methods on two GAN-based speech synthesizers, MelGAN and PWG, but test them on all other synthesizers. Specifically, we split the real \review{speech signals} at a rate of 0.6/0.2/0.2 for training/validation/testing. The fake \review{speech signals} generated by MelGAN and PWG are split at a rate of 0.8/0.2 for training/validation, while those generated by other synthesizers are all used for testing.

Table~\ref{tab:cross-method-LibriSeVoc} reports the cross-method evaluation results on the LibriSeVoc dataset. As can be seen, our method can \review{achieve the best average AUC performance and the second-best average EER performance on the LibriSeVoc dataset}. Though trained on GAN-based synthesizers, our method can still perform relatively well on the other non-GAN synthesizers. To better illustrate the effectiveness of our method, we employ t-SNE~\cite{vandermaaten2008-t-SNE} to analyze the latent features of our method and the comparison methods. Concretely, we run these methods on the validation and test subsets on the LibriSeVoc dataset and collect their latent features, that is, the features before the final classification layer. When using t-SNE~\cite{vandermaaten2008-t-SNE} to cluster these features,
we select 300 samples randomly for the real class and each speech synthesizer. Fig.~\ref{fig:t_SNE_cross_method} illustrates the visualization results of feature clustering for each detection method. It is clear that our method is able to separate the features of those seven types of samples. This better feature separation capability enables our method to achieve higher detection performance on unseen synthesizers.

For the WaveFake dataset, the test synthesizers are nearly all GAN-based methods except WaveGlow. Therefore, nearly all detection methods can obtain better EER performance than the previous performance on the LibriseVoc. \review{As can be seen in Table~\ref{tab:cross-method-WaveFake}}, our method achieves the best EER scores on all the synthesizers, and the EER score is approximately close to 0 on each synthesizer. The cross-method detection results on these two datasets demonstrate the high performance of our method on unseen synthesizer methods in the same dataset.

\begin{table}[!t]
    \centering
    \caption{\review{Details of the ASVspoof2021 DF datasets.}}
    \small
    \label{tab:ASVspoof2021_dataset}
    \color{TableColor}
    \begin{tabular}{lrrrrrrrrrrrr}
    \toprule
    & Train   & Validation     & Test  \\
    \midrule
    No. Synthesizers             & 13 & 13 & 101 \\
    No. Real             &  4795    & 973     & 14869 \\
    No. Fake                 &  44530   & 9027    & 65273  \\
    Total                    &  49325   & 10000   & 80142 \\
    \bottomrule
    \end{tabular}
\end{table}
\begin{table}[!t]
    \color{TableColor}
    \footnotesize
    \centering
    \caption{\review{EER($\downarrow$) ($\%$) performances on the ASVspoof2021 DF test subset. The best scores are formatted to \best{red}, and the second best scores are formatted to \secondbest{violet}.}}
    \label{tab:ASV2021}
    \setlength\tabcolsep{3pt}
    \begin{tabular}{lcrrrrrrrrrr}
    \toprule
    \multirow{2.5}{*}{Method} & \multirow{2.5}{*}{\tabincell{c}{Seen\\ Synthesizers}} & \multicolumn{5}{c}{Unseen Synthesizers} & \multirow{2.5}{*}{\tabincell{c}{Whole\\Testing}} \\ \cmidrule{3-7}
    & & \multicolumn{1}{c}{AR} & \multicolumn{1}{c}{NAR} & \multicolumn{1}{c}{TRD} & \multicolumn{1}{c}{UNK} & \multicolumn{1}{c}{CONC} & \\
    \midrule
LCNN & \secondbest{15.31} & 23.80 & 25.40 & 17.47 & 19.01 & \secondbest{17.16} & 21.30 \\
RawNet2 & 20.34 & 27.95 & 26.62 & 17.17 & 24.57 & 22.39 & 24.31 \\
RawGAT & 14.70 & 23.50 & 18.46 & \best{8.41} & 18.74 & 17.92 & \secondbest{18.07} \\
Wave2Vec2 & 36.82 & 33.92 & 33.22 & 32.85 & 31.89 & 38.03 & 33.60 \\
WaveLM & 25.59 & \secondbest{21.69} & \best{16.62} & 13.05 & \secondbest{15.50} & 27.64 & 18.76 \\
RawNet2-Voc & 18.36 & 28.87 & 27.07 & 14.77 & 23.82 & 19.72 & 24.09 \\
AudioClip & 19.06 & 25.98 & 27.79 & 20.01 & 25.72 & 19.82 & 24.67 \\
Wav2Clip & 14.33 & 25.20 & 27.31 & 15.95 & 19.25 & 12.39 & 20.68 \\
AASIST & 16.05 & 23.97 & 19.48 & \secondbest{9.04} & 19.45 & 19.50 & 19.02 \\
SFATNet & 25.30 & 30.89 & 30.86 & 27.24 & 30.06 & 25.87 & 29.51 \\
ASDG & 19.03 & 27.39 & 27.16 & 21.46 & 24.61 & 20.82 & 25.02 \\
\textbf{Ours} & \best{\phantom{0}8.79} & \best{18.12} & \secondbest{18.31} & 9.06 & \best{13.03} & \best{9.89} & \best{14.79} \\
    \bottomrule
    \end{tabular}
\end{table}

\begin{table*}[t]
    \centering
    \footnotesize
    \caption{Cross-dataset evaluation on the WaveFake dataset \review{and on the EN and ZH subsets of the DECRO dataset}. All the models are trained and validated on the LibriSeVoc dataset but tested on the \review{evaluation datasets}. We report the AUC($\uparrow$) / EER($\downarrow$) ($\%$)  performance.
    }
    \label{tab:cross-dataset}
    \color{TableColor}
    
    \setlength{\tabcolsep}{3pt}{
    \begin{tabular}{lrrrrrrrrrrrrrrrrrr}
    \toprule
    \multirow{2.5}{*}{Method} & \multicolumn{8}{c}{Synthesizers in WaveFake}& \multicolumn{2}{c}{DECRO} \\ \cmidrule(lr){2-9} \cmidrule(l){10-11}
     &  \multicolumn{1}{c}{MelGAN} &  \multicolumn{1}{c}{PWG} &  \multicolumn{1}{c}{MB-MelGAN} &  \multicolumn{1}{c}{FB-MelGAN} &  \multicolumn{1}{c}{HiFi-GAN} &  \multicolumn{1}{c}{MelGAN-L} &  \multicolumn{1}{c}{WaveGLow}&  \multicolumn{1}{c}{\textit{Average}} & \multicolumn{1}{c}{ZH}&  \multicolumn{1}{c}{EN} \\
    \midrule
LCNN & 99.96/\phantom{0}0.75 & \secondbest{95.07}/\secondbest{11.95} & 97.02/\phantom{0}9.03 & \secondbest{97.42}/\secondbest{\phantom{0}8.27} & 97.01/\phantom{0}9.04 & \best{99.98}/\best{\phantom{0}0.50} & 99.97/\phantom{0}0.71 & \secondbest{98.06}/\secondbest{\phantom{0}5.75} & 74.77/30.69 & 61.88/41.42  \\
RawNet2 & 99.63/\phantom{0}2.80 & 80.31/27.40 & 87.83/20.36 & 66.89/37.99 & 78.15/29.40 & 97.39/\phantom{0}8.27 & 99.97/\phantom{0}0.70 & 87.17/18.13 & 61.37/41.52 & 50.02/50.71  \\
RawGAT & 99.99/\phantom{0}0.22 & 90.13/17.74 & 97.19/\phantom{0}8.42 & 87.14/20.92 & 87.02/20.99 & 99.95/\secondbest{\phantom{0}0.57} & 99.96/\phantom{0}0.70 & 94.48/\phantom{0}9.94 & 71.08/34.70 & 64.38/39.86  \\
Wave2Vec2 & 78.83/27.84 & 66.28/38.50 & 80.37/26.11 & 63.10/41.00 & 63.96/40.42 & 72.38/33.76 & 79.90/26.57 & 72.12/33.46 & 81.03/26.17 & 68.20/37.26  \\
WaveLM & 95.59/\phantom{0}9.64 & 90.34/16.69 & 96.95/\phantom{0}7.65 & 88.44/18.78 & 83.92/23.34 & 94.03/12.00 & 86.59/20.33 & 90.83/15.49 & 75.47/28.74 & 59.58/43.89  \\
RawNet2-Voc & 98.20/\phantom{0}6.26 & 71.01/34.28 & 80.10/26.49 & 59.98/42.66 & 71.81/33.74 & 91.07/15.84 & 99.94/\phantom{0}0.97 & 81.73/22.89 & 62.33/40.87 & 45.92/53.34  \\
AudioClip & 98.64/\phantom{0}5.19 & 87.38/19.64 & 97.56/\phantom{0}6.89 & 91.92/14.56 & 94.42/11.52 & 96.42/\phantom{0}8.82 & 99.54/\phantom{0}2.72 & 95.13/\phantom{0}9.91 & 70.00/35.21 & \secondbest{73.39}/\secondbest{31.59}  \\
Wav2Clip & 99.11/\phantom{0}3.36 & 93.41/13.52 & \secondbest{98.03}/\secondbest{\phantom{0}6.57} & 96.10/\phantom{0}9.99 & \secondbest{97.02}/\secondbest{\phantom{0}8.57} & 97.63/\phantom{0}6.02 & 99.96/\phantom{0}0.66 & 97.32/\phantom{0}6.96 & \secondbest{83.70}/\secondbest{23.80} & 67.63/38.34  \\
AASIST & \best{100.00}/\secondbest{\phantom{0}0.18} & 91.92/16.03 & 97.70/\phantom{0}7.74 & 87.29/20.89 & 88.30/19.89 & 99.97/\phantom{0}0.61 & \secondbest{99.98}/\secondbest{\phantom{0}0.51} & 95.02/\phantom{0}9.41 & 71.13/34.12 & 70.44/35.57  \\
SFATNet & 92.08/16.19 & 90.21/18.46 & 80.07/27.83 & 69.98/36.06 & 73.50/33.01 & 86.97/21.32 & 98.92/\phantom{0}5.02 & 84.53/22.56 & 66.03/36.14 & 67.93/36.50  \\
ASDG & 99.85/\phantom{0}1.86 & 84.40/23.38 & 92.58/14.93 & 94.94/11.69 & 93.94/13.22 & 99.99/\phantom{0}0.66 & 99.97/\phantom{0}0.88 & 95.10/\phantom{0}9.52 & 72.53/32.03 & 54.30/49.00  \\
\textbf{Ours} & \best{100.00}/\best{\phantom{0}0.09} & \best{99.35}/\best{\phantom{0}3.86} & \best{99.41}/\best{\phantom{0}3.57} & \best{99.41}/\best{\phantom{0}3.62} & \best{99.35}/\best{\phantom{0}3.80} & \best{99.99}/\best{\phantom{0}0.27} & \best{100.00}/\best{\phantom{0}0.04} & \best{99.64}/\best{\phantom{0}2.18} & \best{88.84}/\best{17.77} & \best{97.81}/\best{\phantom{0}6.88} \\
    \bottomrule
    \end{tabular}
    }
\end{table*}

\subsubsection{\review{ASVSpoof2021}}
\review{We conduct experiments on the ASVSpoof2021 Deepfake (DF) dataset for further cross-method evaluation, since it provides standard splits and contains synthetic generation methods that are not shared between training, validation, and testing.} 

\review{The split details of this dataset are shown in Table~\ref{tab:ASVspoof2021_dataset}. In this dataset, the synthesizer methods are divided into five categories: neural vocoder autoregressive (AR), neural vocoder non-autoregressive (NAR), traditional vocoder (TRD), waveform concatenation (CONC), and unknown (UNK). It should be noted that we use only a portion of fake samples in the test subset of the ASVspoof2021 DF dataset. Specifically, the number of fake samples in each synthesizer category matches the number of real samples in the test subset. The comparison results on the ASVSpoof2021 DF dataset are presented in Table~\ref{tab:ASV2021}. As can be seen, our method achieves better detection performance on both seen and unseen synthesizers.}

\subsection{Cross-Dataset Evaluation}

In the cross-dataset evaluation, we train and validate all the methods on the LibriSeVoc~\cite{sunAISynthesizedVoiceDetection2023-LibraSeVoc} dataset and test them on the WaveFake~\cite{frank2021wavefake} and DECRO~\cite{ba2023transferring-DECRO} datasets. Specifically, we split the whole LibriSeVoc dataset at 0.8/0.2 for training/validation. For each synthesizer of the WaveFake dataset, we combine its generated fake \review{speech samples} and all the real \review{speech samples} to build a subset for testing. As for the evaluation of the DECRO dataset, we test all the detection methods on the test splits of the EN and ZH subsets, respectively.

\begin{figure*}
    \centering
    \small

    \begin{minipage}[b]{0.95\linewidth}
        \centerline{\includegraphics[width=1\linewidth]{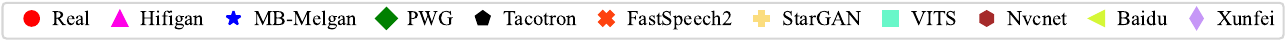}}
    \end{minipage}

    \begin{minipage}[b]{0.99\linewidth}
    \centering
        \begin{minipage}[b]{0.16\linewidth}
            \centerline{\includegraphics[width=1\linewidth]{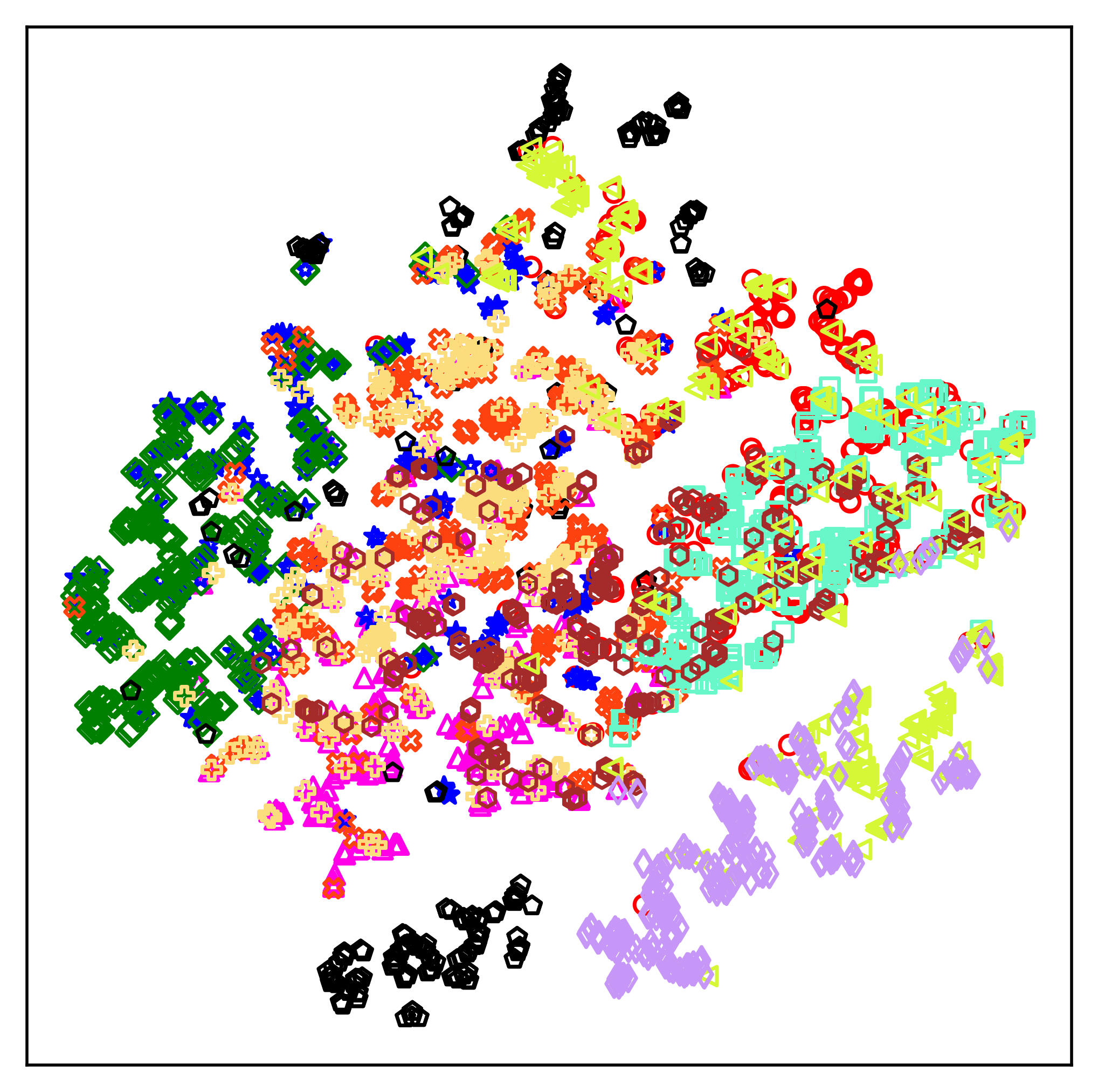}}
            \centerline{LCNN~\cite{lavrentyeva2019stc-LCNN}}
        \end{minipage}
        \begin{minipage}[b]{0.16\linewidth}
            \centerline{\includegraphics[width=1\linewidth]{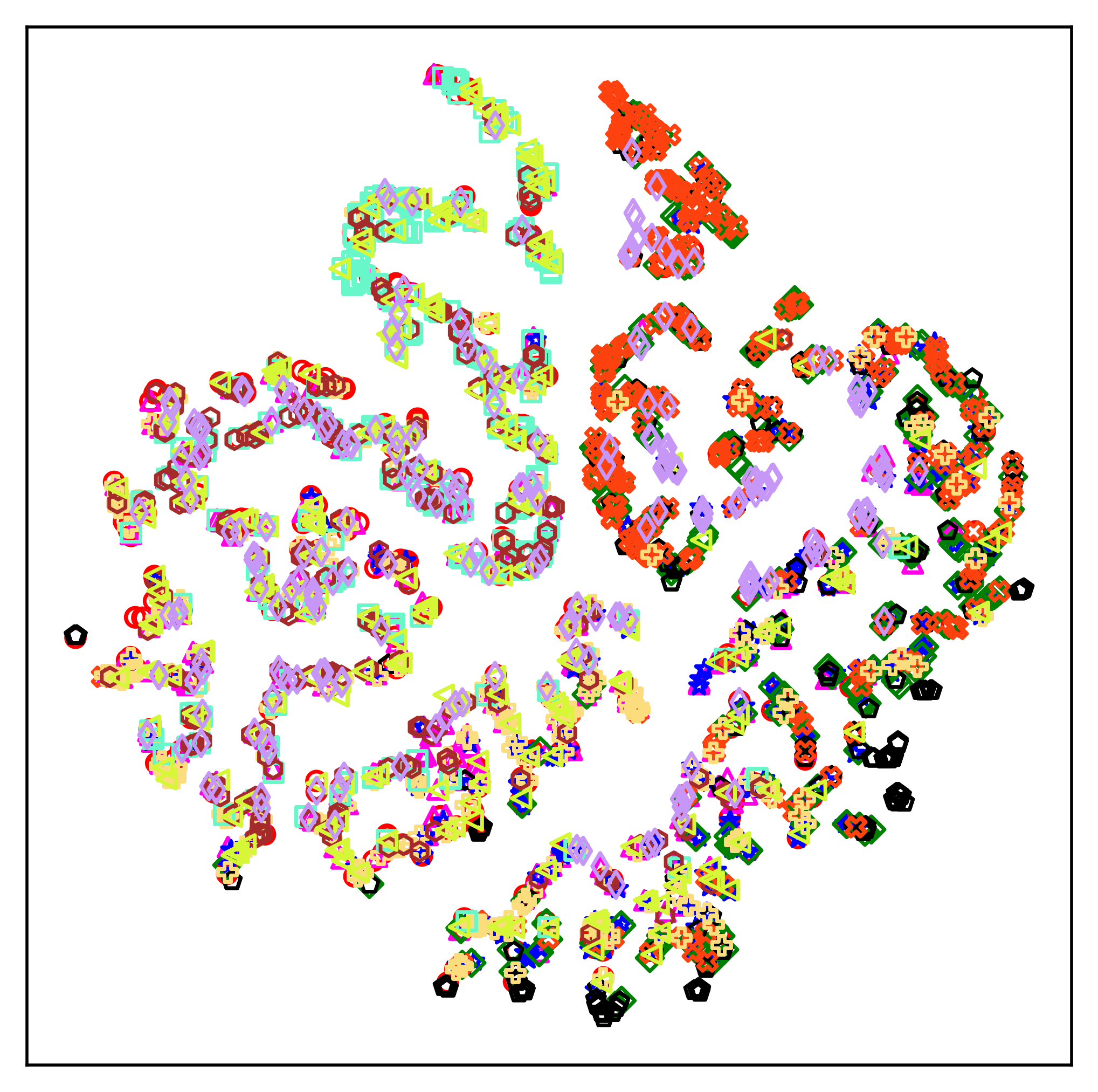}}
            \centerline{RawNet2~\cite{jung2020improved-RawNet2}}
        \end{minipage}
        \begin{minipage}[b]{0.16\linewidth}
            \centerline{\includegraphics[width=1\linewidth]{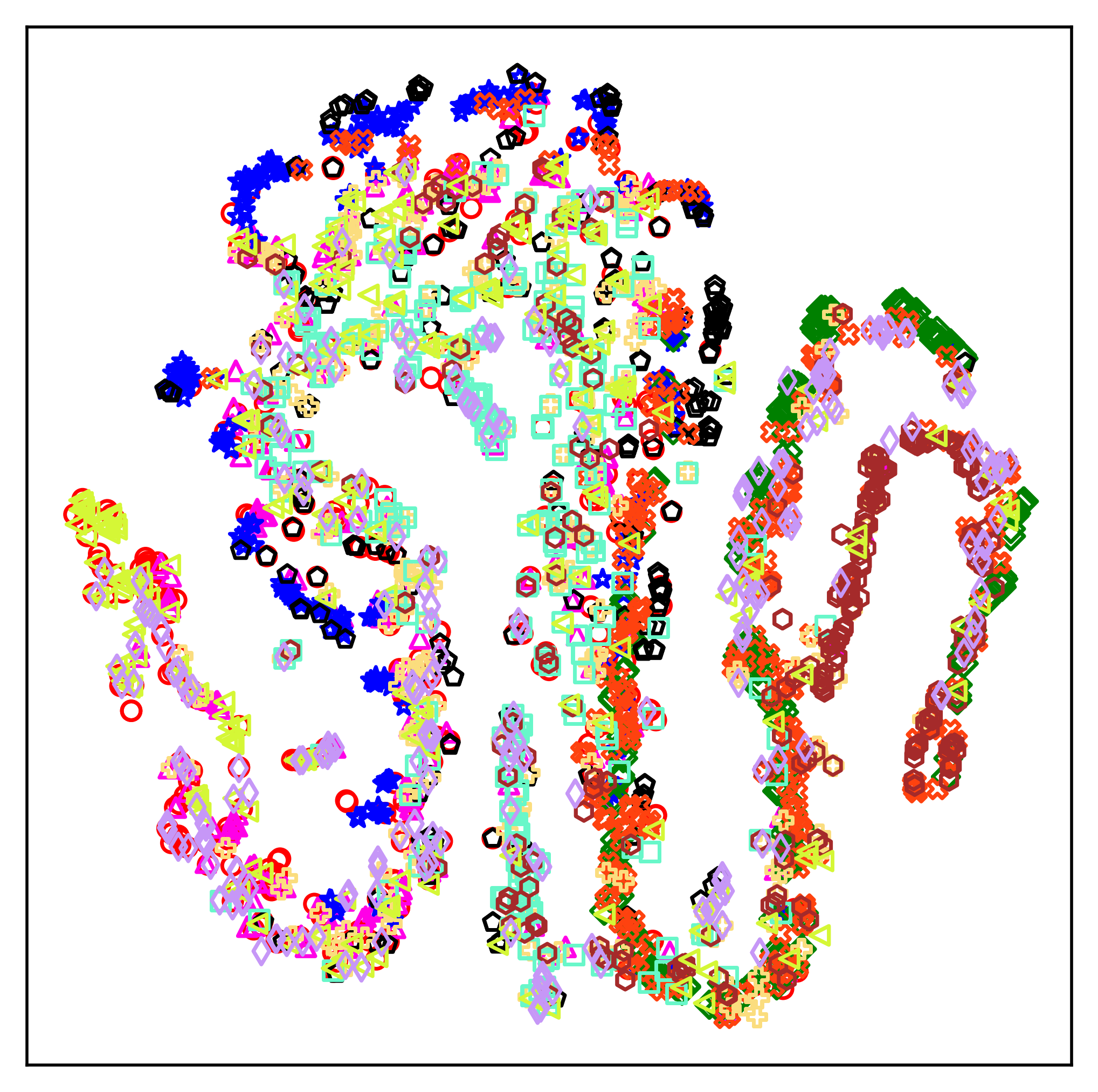}}
            \centerline{RawGAT~\cite{tak21_asvspoof_RawGAT}}
        \end{minipage}
        \begin{minipage}[b]{0.16\linewidth}
            \centerline{\includegraphics[width=1\linewidth]{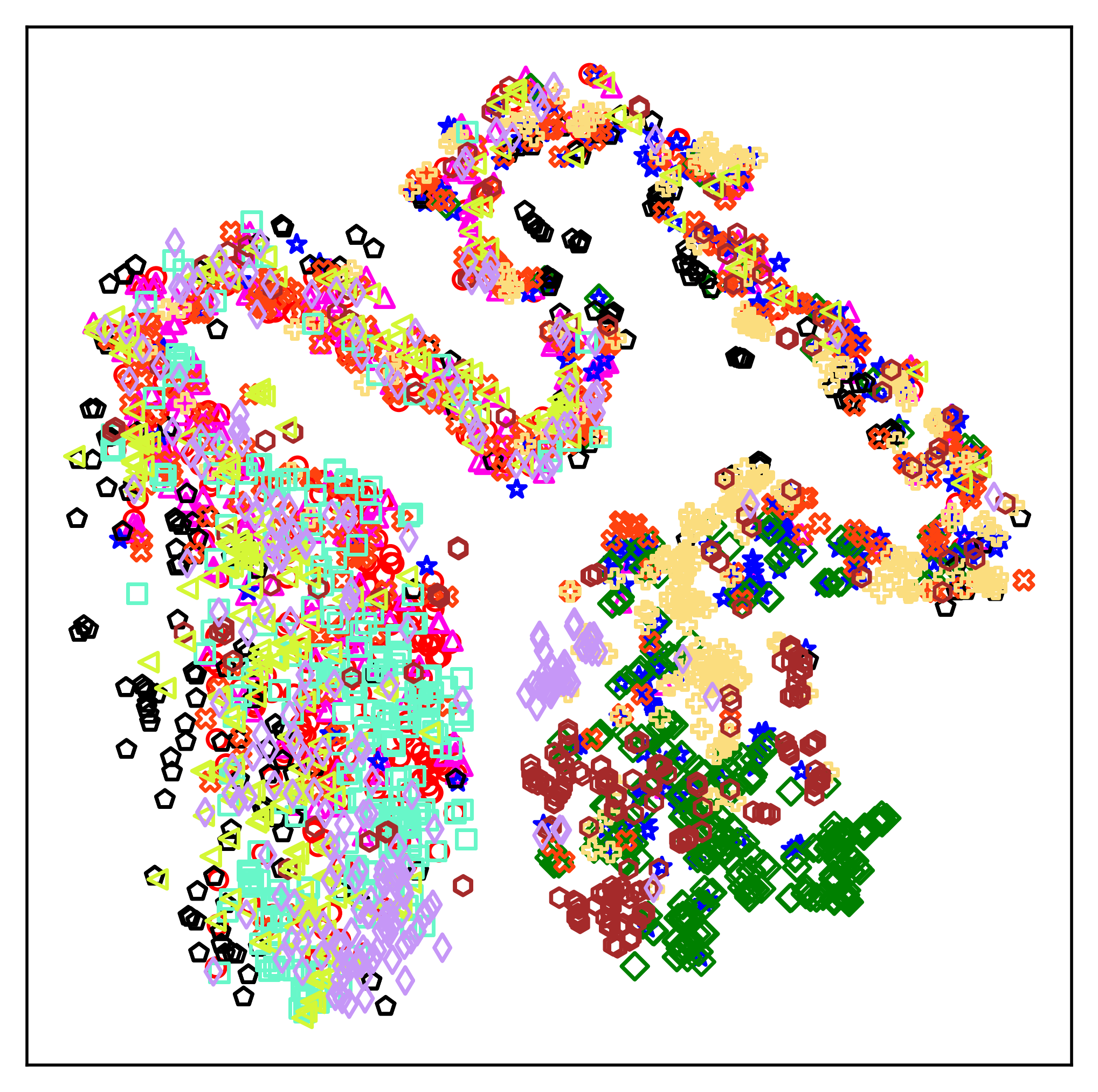}}
            \centerline{Wave2Vec2~\cite{baevski2020wav2vec}}
        \end{minipage}
        \begin{minipage}[b]{0.16\linewidth}
            \centerline{\includegraphics[width=1\linewidth]{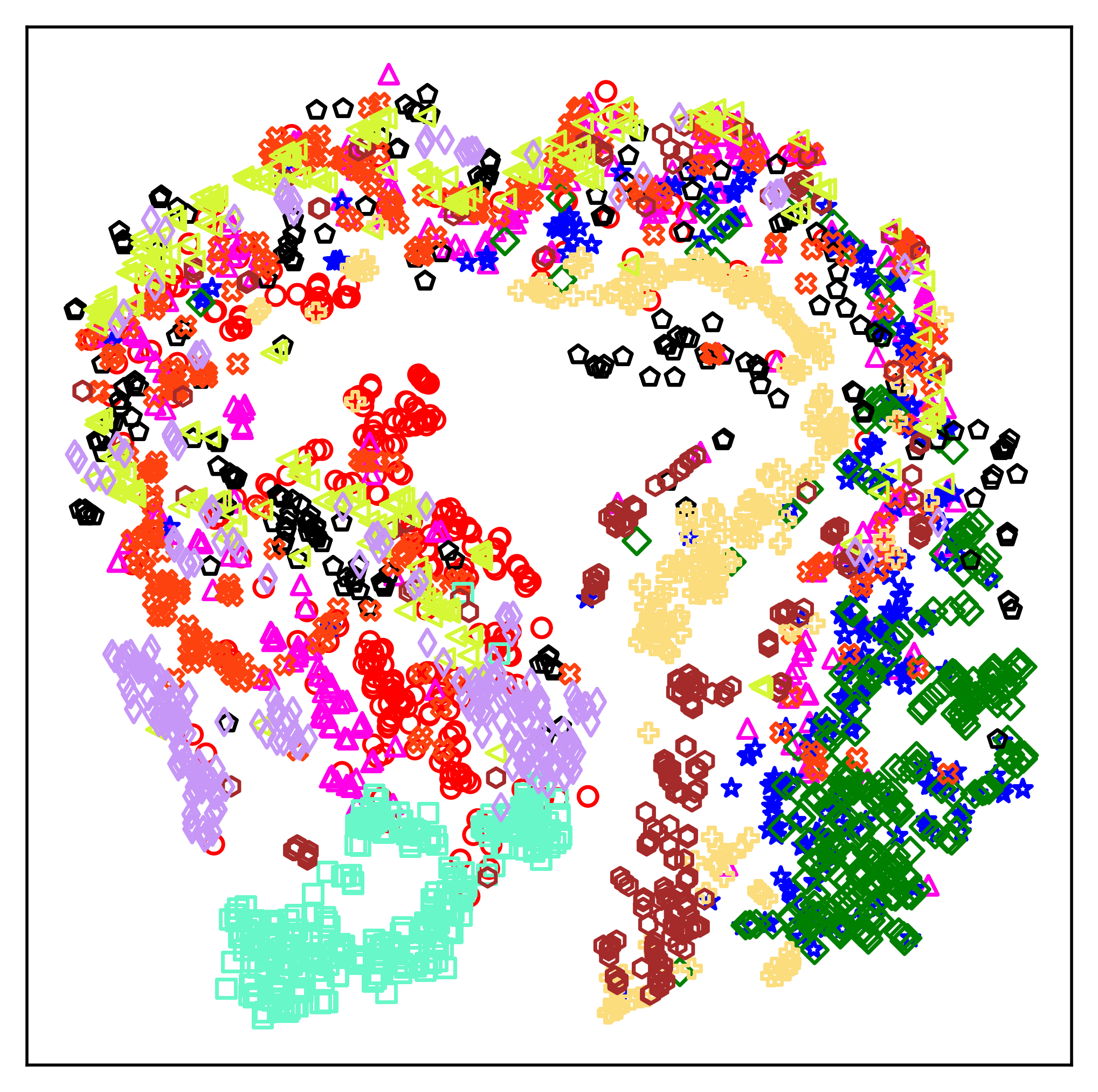}}
            \centerline{WaveLM~\cite{chen2022wavlm}}
        \end{minipage}
        \begin{minipage}[b]{0.16\linewidth}
            \centerline{\includegraphics[width=1\linewidth]{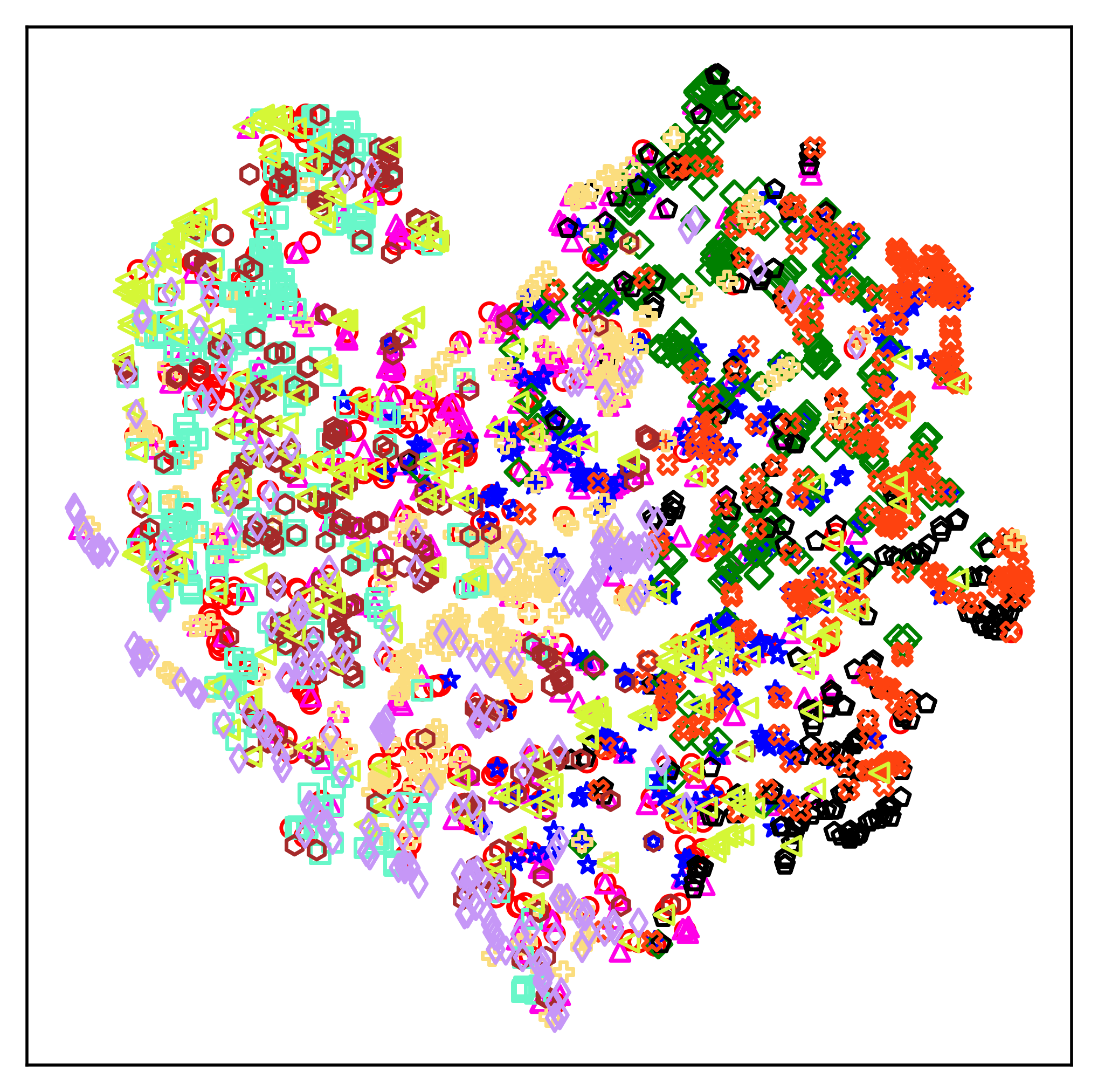}}
            \centerline{RawNet2-Voc~\cite{sunAISynthesizedVoiceDetection2023-LibraSeVoc}}
        \end{minipage}
        
        \vspace{5pt}

        \begin{minipage}[b]{0.16\linewidth}
            \centerline{\includegraphics[width=1\linewidth]{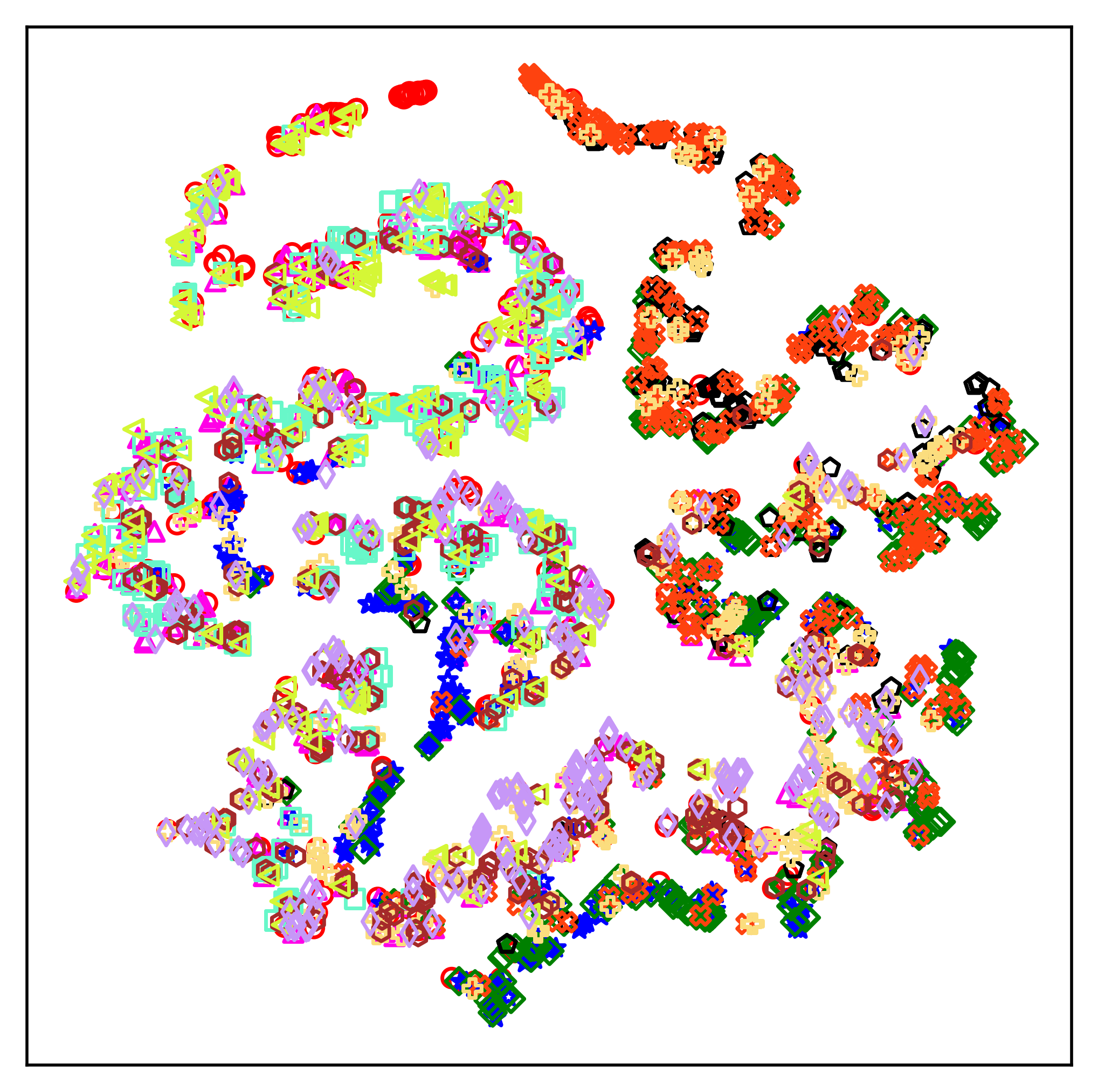}}
            \centerline{AudioClip~\cite{guzhov2022audioclip}}
        \end{minipage}
        \begin{minipage}[b]{0.16\linewidth}
            \centerline{\includegraphics[width=1\linewidth]{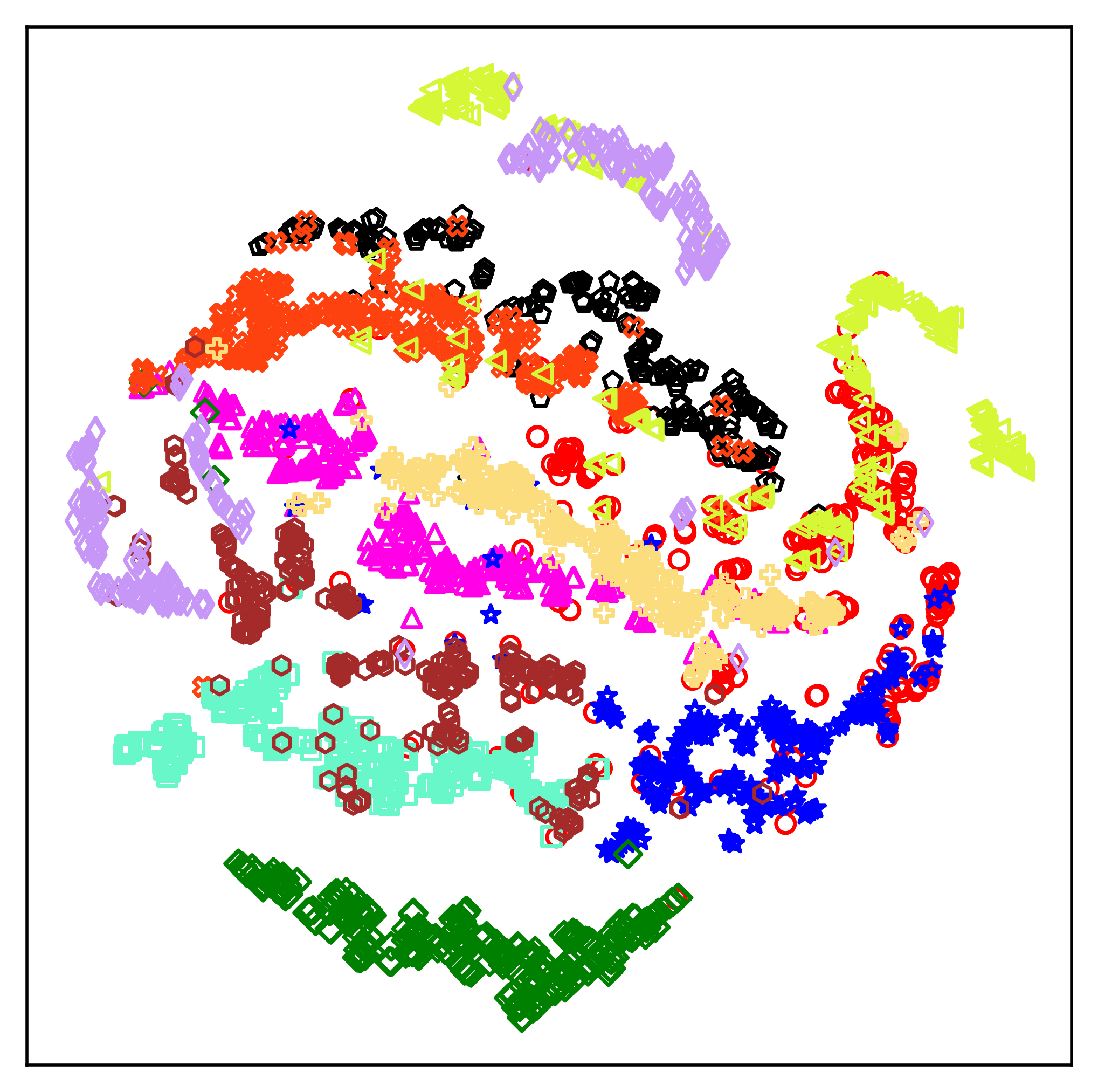}}
            \centerline{Wav2Clip~\cite{wu2022wav2clip}}
        \end{minipage}
        \begin{minipage}[b]{0.16\linewidth}
            \centerline{\includegraphics[width=1\linewidth]{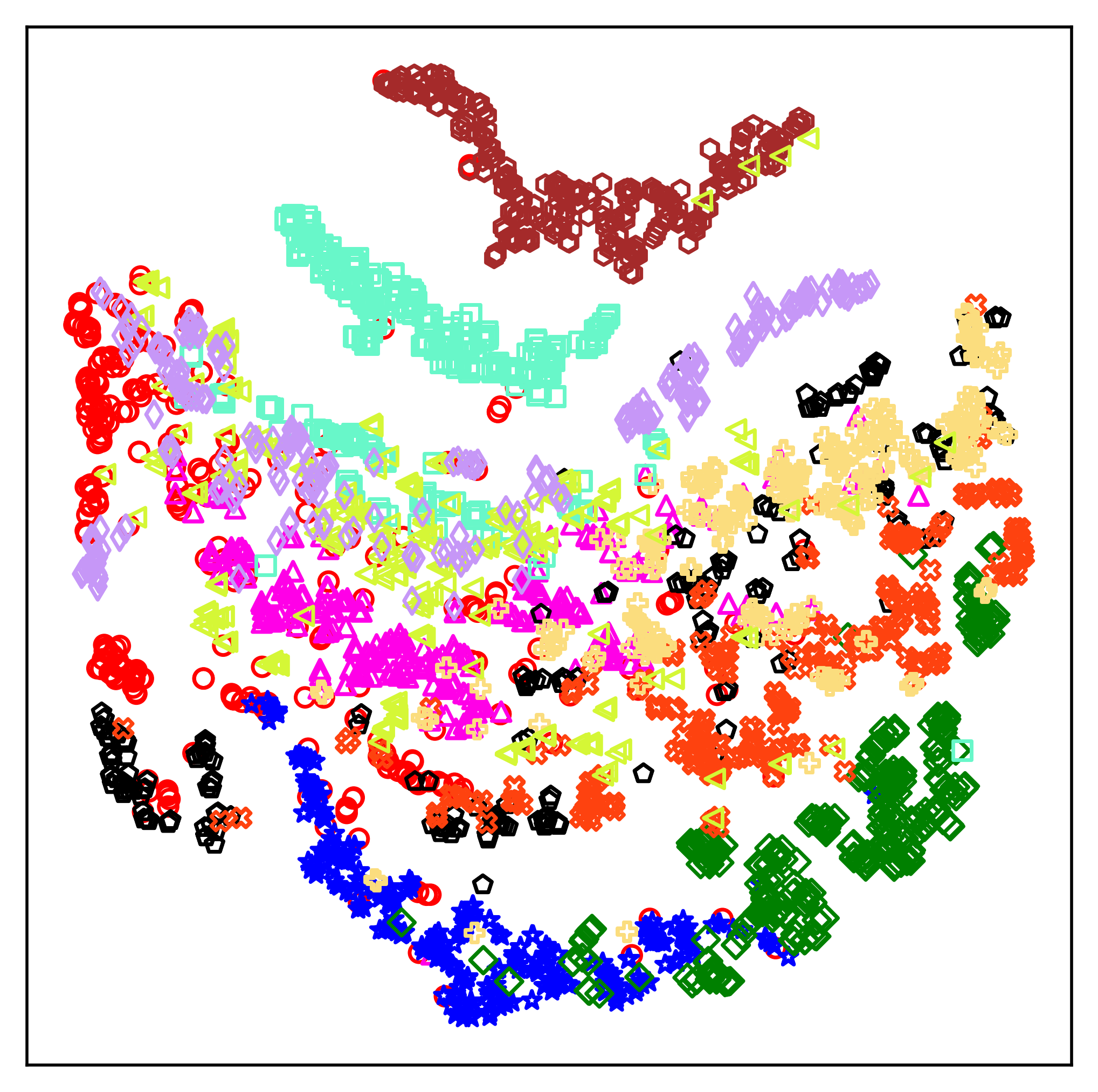}}
            \centerline{AASIST~\cite{jung2022aasist}}
        \end{minipage}
        \begin{minipage}[b]{0.16\linewidth}
            \centerline{\includegraphics[width=1\linewidth]{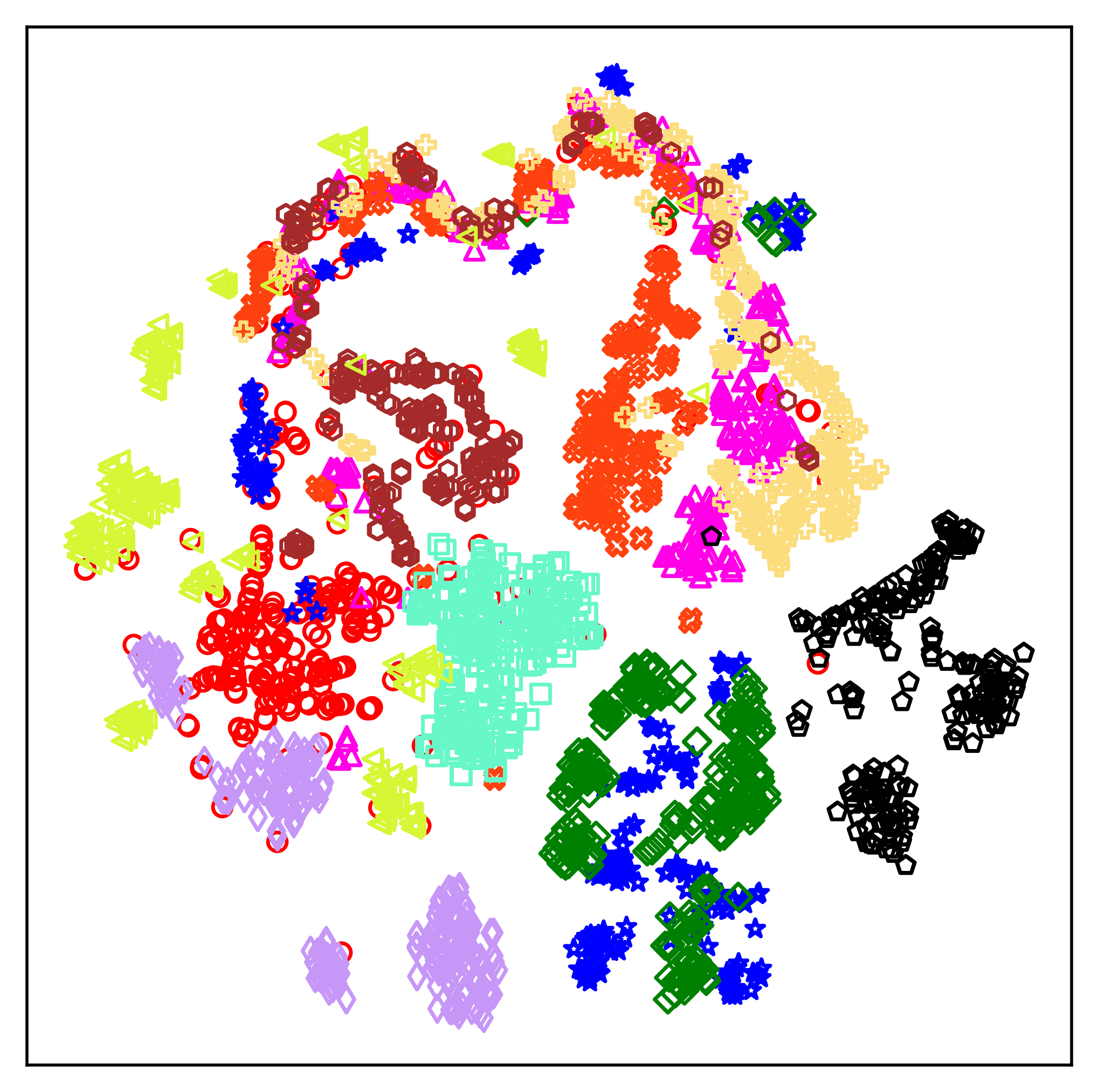}}
            \centerline{\review{ASDG}~\cite{xie2024domain_ASDG}}
        \end{minipage}
        \begin{minipage}[b]{0.16\linewidth}
            \centerline{\includegraphics[width=1\linewidth]{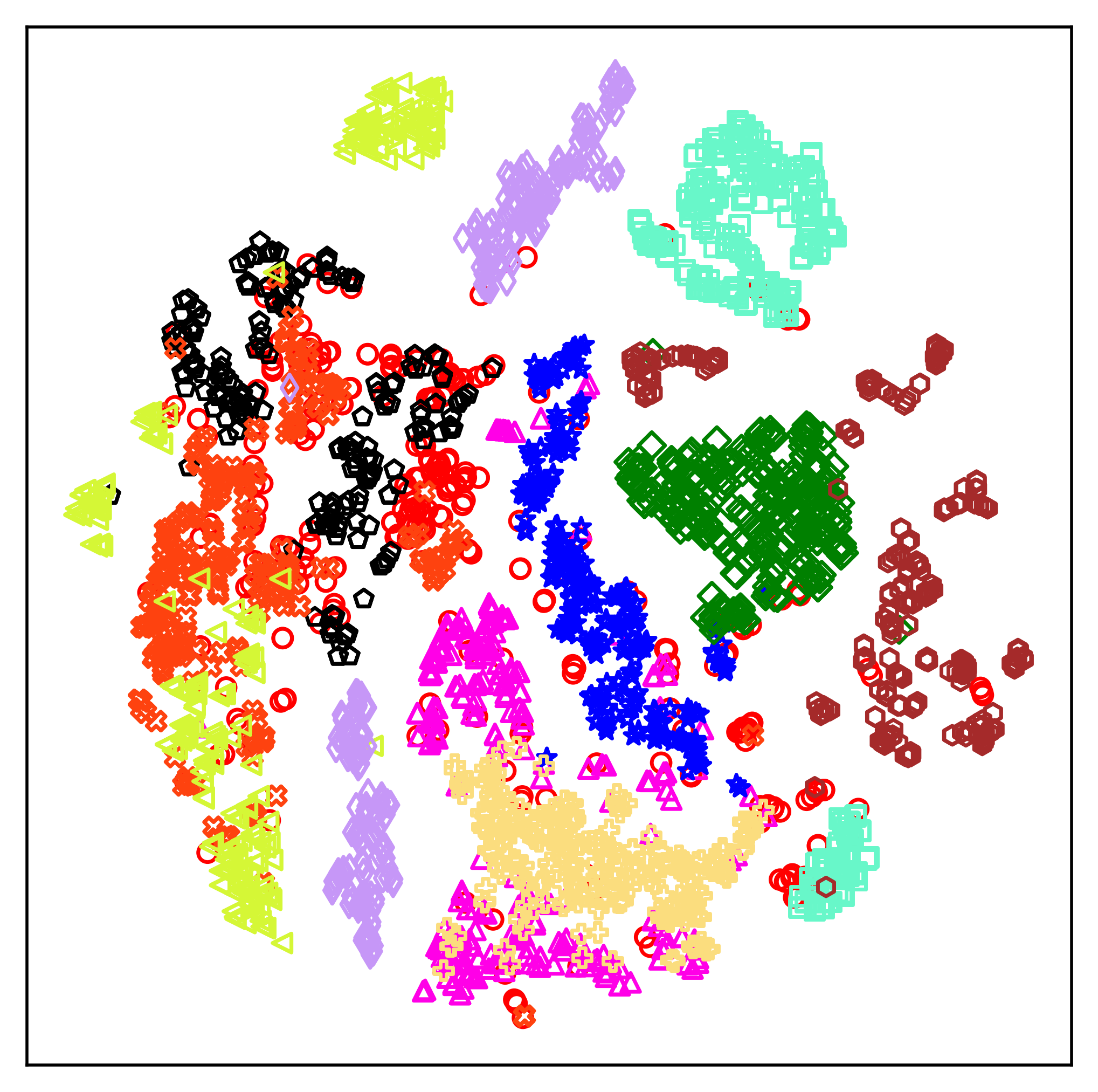}}
            \centerline{\review{SFATNet}~\cite{cuccovillo_audio_2023_f0}}
        \end{minipage}
        \begin{minipage}[b]{0.16\linewidth}
            \centerline{\includegraphics[width=1\linewidth]{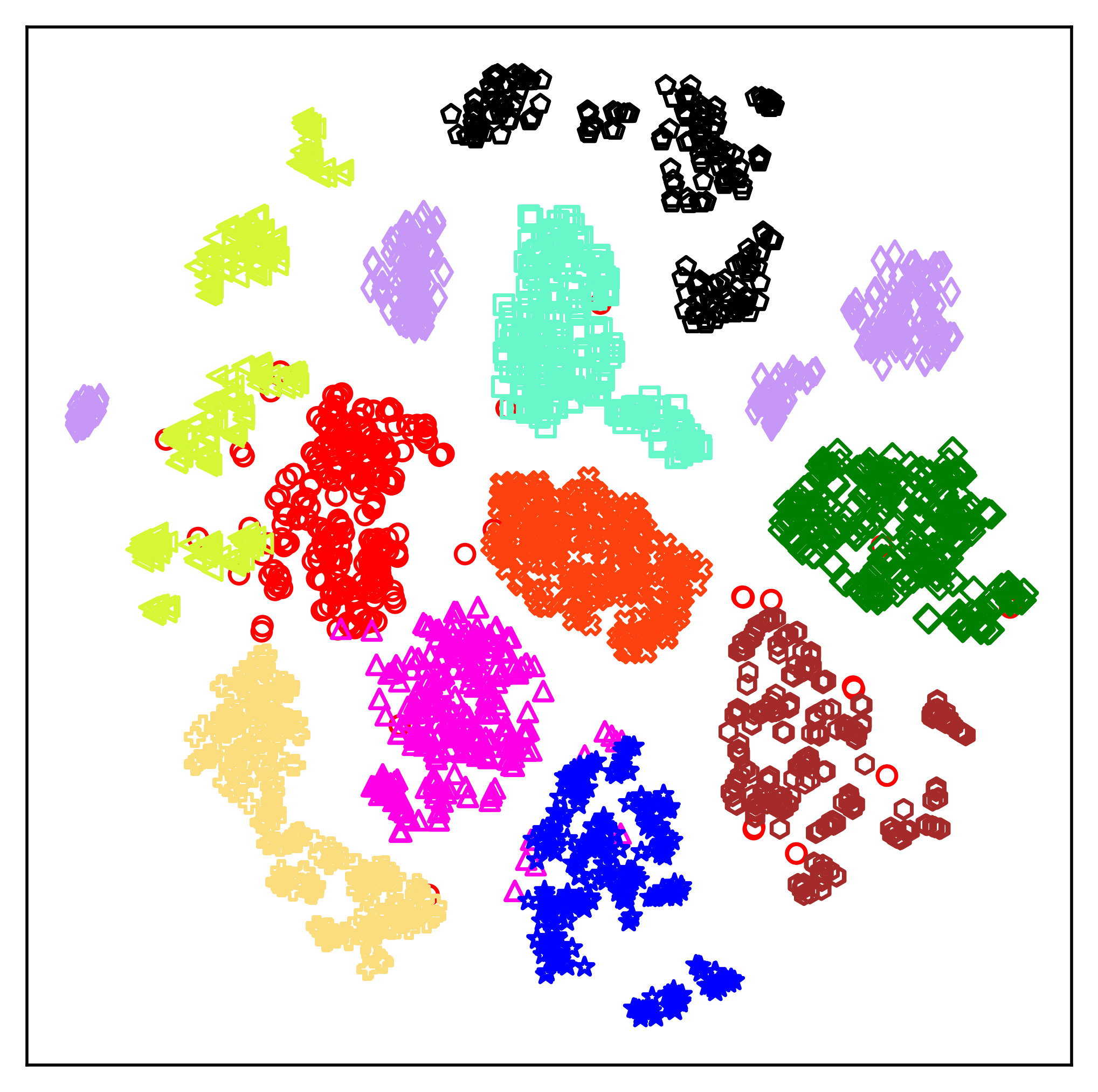}}
            \centerline{\textbf{Ours}}
        \end{minipage}
    \end{minipage}

    \caption{T-SNE visualization in the cross-evaluation task on the DECRO ZH subset. For each deepfake speech detection method, we extract the latent features from the validation and test subsets and randomly extract 300 samples of the real and each fake method for visualization.}
    \label{fig:t_SNE_cross_dataset}
\end{figure*}

Table~\ref{tab:cross-dataset} illustrates the cross-dataset results on the WaveFake dataset \review{and on the EN and ZH subsets of the DECRO dataset}. One can see that our method achieves the best EER \review{on all the synthesizers and obtains the best average EER of 2.18\% on the evaluation}. 
As seen \review{from the last two columns, our method can achieve the EER of 6.88\% and 17.77\% in the EN and ZH subsets of the DECRO dataset}, respectively, and has better performance than the comparison methods. To better illustrate the effectiveness of our method, we employ t-SNE~\cite{vandermaaten2008-t-SNE} to analyze the latent features of our method and the comparison methods on the ZH subset of the DECRO dataset. The visualization results are shown in Fig.~\ref{fig:t_SNE_cross_dataset}, where our method can effectively separate the eleven types of features. The better clustering effect demonstrates that our method can learn more discriminative features.

These evaluation results in Table~\ref{tab:cross-dataset} and Fig.~\ref{fig:t_SNE_cross_dataset} demonstrate that our method surpasses existing detection methodologies by achieving notably lower EER scores across diverse datasets. This achievement underscores the robustness and effectiveness of our method in discerning deepfake audio signals across different sources and speech forgery methods.

\subsection{Cross-Language Evaluation}


\begin{table}[t]
    \centering
    \caption{Cross-lingual evaluation on the WaveFake and DECRO datasets. ``A$\rightarrow$B'' means that models are trained and validated on the language A but tested on the language B. We report the EER ($\%$) performance, and the best scores are set to \textbf{bold}.}
    \label{tab:cross_language}
    \small
    \color{TableColor}
    
    \begin{tabular}{lrrrrrrrrrrrr}
    \toprule
    \multirow{2}{*}{Model} & WaveFake & \multicolumn{2}{c}{DECRO} & \multirow{2}{*}{\textbf{\textit{Average}}}\\  \cmidrule(lr){2-2} \cmidrule(l){3-4}
      & EN$\rightarrow$JP & ZH$\rightarrow$EN & EN$\rightarrow$ZH & \\
    \midrule
LCNN & \best{6.74} & 44.14 & 35.68 & 28.85 \\
RawNet2 & 32.93 & 44.97 & 45.17 & 41.02 \\
RawGAT & 12.86 & 43.66 & 42.53 & 33.02 \\
Wave2Vec2 & 16.90 & 37.10 & 30.49 & 28.16 \\
WaveLM & 10.22 & 42.37 & 35.55 & 29.38 \\
RawNet2-Voc & 36.80 & 35.81 & 41.74 & 38.11 \\
AudioClip & 46.66 & 49.45 & 45.26 & 47.12 \\
Wav2Clip & 23.17 & 33.74 & \best{21.66} & \secondbest{26.19} \\
AASIST & 13.04 & 43.64 & 40.90 & 32.53 \\
SFATNet & 37.03 & \secondbest{30.47} & 33.79 & 33.76 \\
ASDG & \secondbest{7.25} & 43.30 & 36.04 & 28.86 \\
\rowcolor{LightCyan}
\textbf{Ours} & 23.26 & \best{16.66} & \secondbest{27.54} & \best{22.48} \\
    \bottomrule
    \end{tabular}
\end{table}

As shown in Table~\ref{tab:datasets}, WaveFake and DECRO datasets contain \review{speech samples} in two languages. We conduct experiments on them to evaluate the cross-lingual ability of all detection methods. For the WaveFake dataset, we train and validate detection methods in the English subset at a split rate of 0.8/0.2 and test them in all Japanese \review{speech samples}. Note that the deepfake \review{speech samples} used in training are only generated by PWG and MB-MelGAN since the Japanese subset only uses these two synthesizers for speech generation. For the DECRO dataset, we conduct two experiments, ``ZH$\rightarrow$EN" and ``EN$\rightarrow$ZH", where the model is trained and validated on the training and validation subsets of one language but tested on the test sub-part of the other language. 

Table~\ref{tab:cross_language} shows the cross-lingual evaluation results. As can be seen, each model has a different level of EER performance across the evaluation tasks. The LCNN method has the lowest EER score when tested with the WaveFake dataset, while our method has the lowest EER scores on the ``ZH$\rightarrow$EN" task in the DECRO dataset. Overall, our method performs comparatively better, with \review{the best overall average EER score 22.48\% among all the models}. This indicates the better generalization ability of our method for cross-lingual tasks.

\begin{table}[!t]
    \centering
    \color{TableColor}
    \small
    \caption{\review{Time complexity and throughput (samples per second) of different detection methods.}}
    \label{tab:complexity}
    \setlength\tabcolsep{3pt}
    \begin{tabular}{lrrrr}
        \toprule
        Method & Parameters & FLOPs & \tabincell{r}{Training\\ Throughput} & \tabincell{r}{Testing\\ Throughput} \\
        \midrule
        LCNN & 0.68 M & 0.25 G & 503.52 & 1734.57 \\
        RawNet2 & 17.70 M & 1.19 G & 562.00 & 1332.46 \\
        RawGAT & 0.44 M & 13.68 G & 89.33 & 258.34 \\
        Wave2Vec2 & 94.40 M & 21.18 G & 281.74 & 926.79 \\
        WaveLM & 94.40 M & 20.79 G & 247.98 & 700.59 \\
        RawNet2-Voc & 17.70 M & 1.19 G & 485.36 & 1525.98 \\
        AudioClip & 134.00 M & 3.26 G & 557.17 & 1942.10 \\
        Wav2Clip & 11.70 M & 1.34 G & 754.44 & 2048.39 \\
        AASIST & 0.30 M & 7.10 G & 158.95 & 461.36 \\
        SFATNet & 81.40 M & 16.30 G & 364.64 & 923.17 \\
        ASDG & 1.10 M & 0.34 G & 1005.62 & 1213.25 \\
        \textbf{Ours} & 22.50 M & 3.21 G & 233.29 & 1841.33 \\
        \bottomrule
    \end{tabular}
\end{table}

\subsection{\review{Time complexity}}
\review{In this section, we analyze the model’s complexity regarding the number of parameters and Floating Point Operations per second (FLOPs). We also report training and testing throughputs under identical hardware conditions with a batch size of 32. The comparison results are presented in Table~\ref{tab:complexity}. As shown, the computational overhead and throughput of our method are comparable to most existing methods. However, our method achieves significantly better detection performance than the comparison methods, as shown in previous comparisons in Tables II-VIII.
}

\begin{table}[t]
    \centering
    \small
    \caption{Ablation results (EER (\%)) of loss functions on two tasks.}
    \label{tab:betas}
    \begin{tabular}{cccccrr}
    \toprule
    Setting  & $\beta_0$ & $\beta_1$ & $\beta_2$ & $\beta_3$ & Task1 & Task2 \\
    \midrule

    (a) & 0.5 & 0.5 & 0.5 & 0.5  & \review{7.13} & \review{2.50}\\ 
    (b) & 1.0 & 1.0 & 0.5 & 0.5  & \review{7.62} & \review{3.09}\\ 
    (c) & 1.0 & 0.5 & 1.0 & 0.5  & \review{8.28} & \review{2.87}\\ 
    (d) & 1.0 & 0.5 & 0.5 & 1.0  & \review{7.02} & \review{2.23}\\ 
    Single-Stream & 0   & 0   & 0   & 0.5  & \review{11.40} & \review{4.63} \\ 
   \review{No synthesizer stream} & \review{1.0} & \review{0} & \review{0.5} & \review{0.5}  & \review{9.01} & \review{3.03} \\
   \review{No content stream} & \review{1.0} & \review{0.5} & \review{0} & \review{0.5}  &  \review{11.81} & \review{4.69}\\
    \rowstyle{\color{black}}
    Default & 1.0 & 0.5 & 0.5 & 0.5  &\review{\textbf{6.12}} & \review{\textbf{2.18}}\\ 
     \bottomrule
    \end{tabular}
\end{table}

\section{Ablation Study and Discussion}
\label{sec:ablation}
In this section, we conduct ablation studies to evaluate the effectiveness of some components of our method. We report the average EER (\%) performance on the cross-method evaluation task in the LibriSeVoc dataset (Task1) and the average EER (\%) performance on the WaveFake dataset in the cross-dataset evaluation (Task2) for all settings.

\subsection{Feature Decomposition Strategy }

\begin{figure*}[!t]
    \centering
    \small

    \begin{minipage}[b]{0.99\linewidth}
    \centering
        \begin{minipage}[b]{0.03\linewidth}
            \centerline{\rotatebox{90}{Log-scale Spectrogram}}
            \vspace{30pt}
            \centerline{\rotatebox{90}{$\mathbf{F}_s$-based CAM}}
            \vspace{30pt}
            \centerline{\rotatebox{90}{$\mathbf{F}_c$-based CAM}}
            \vspace{20pt}
        \end{minipage}
        \begin{minipage}[b]{0.18\linewidth}
            \centerline{\includegraphics[width=1\linewidth]{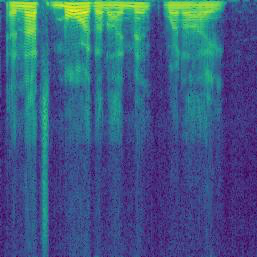}}
            \vspace{2pt}
            \centerline{\includegraphics[width=1\linewidth]{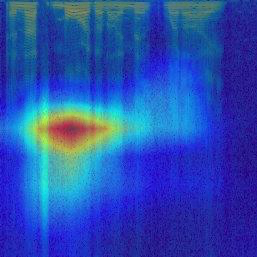}}
            \vspace{2pt}
            \centerline{\includegraphics[width=1\linewidth]{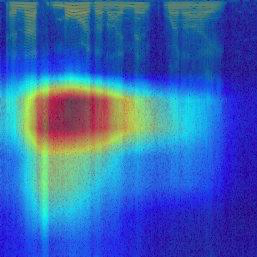}}
            \centerline{\review{\footnotesize{196\_122159\_11\_01\_gen}}}
        \end{minipage}
        \begin{minipage}[b]{0.18\linewidth}
            \centerline{\includegraphics[width=1\linewidth]{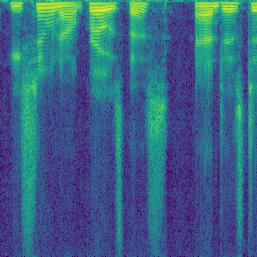}}
            \vspace{2pt}
            \centerline{\includegraphics[width=1\linewidth]{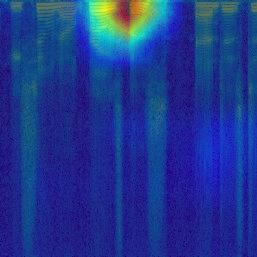}}
            \vspace{2pt}
            \centerline{\includegraphics[width=1\linewidth]{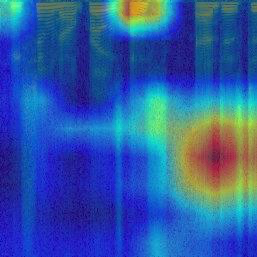}}
            \centerline{\review{\footnotesize{6415\_111615\_03\_01\_gen}}}
        \end{minipage}
        \begin{minipage}[b]{0.18\linewidth}
            \centerline{\includegraphics[width=1\linewidth]{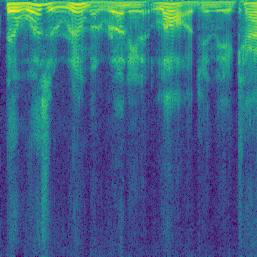}}
            \vspace{2pt}
            \centerline{\includegraphics[width=1\linewidth]{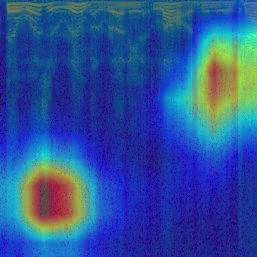}}
            \vspace{2pt}
            \centerline{\includegraphics[width=1\linewidth]{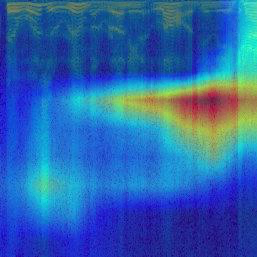}}
            \centerline{\review{\footnotesize{6367\_74004\_04\_10\_gen}}}
        \end{minipage}
        \begin{minipage}[b]{0.18\linewidth}
            \centerline{\includegraphics[width=1\linewidth]{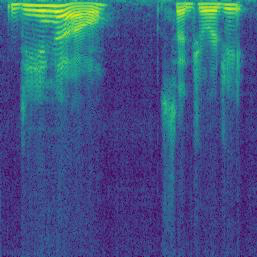}}
            \vspace{2pt}
            \centerline{\includegraphics[width=1\linewidth]{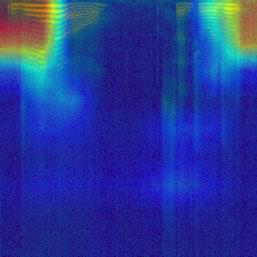}}
            \vspace{2pt}
            \centerline{\includegraphics[width=1\linewidth]{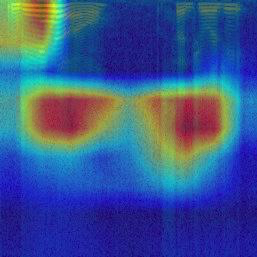}}
            \centerline{\review{\footnotesize{6415\_100596\_60\_00\_gen}}}
        \end{minipage}
        \begin{minipage}[b]{0.18\linewidth}
            \centerline{\includegraphics[width=1\linewidth]{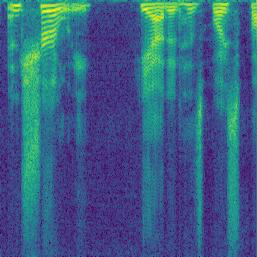}}
            \vspace{2pt}
            \centerline{\includegraphics[width=1\linewidth]{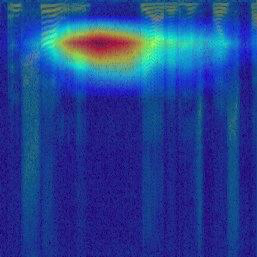}}
            \vspace{2pt}
            \centerline{\includegraphics[width=1\linewidth]{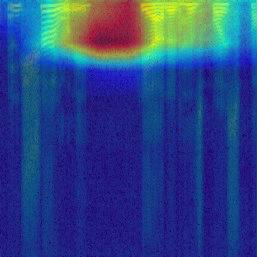}}
            \centerline{\review{\footnotesize{4195\_186237\_10\_01\_gen}}}
        \end{minipage}
    \end{minipage}

    \caption{Grad-CAM visualization on the LibriseVoc dataset. From top to bottom of each column, the three images are raw log-scale spectrograms, gradient visualizations from the synthesizer, and content features. The brighter color denotes a larger influence on the classification. \review{Note all the used speech samples are deepfake from the DiffWave subset, and the column names denote the file names.}
    }
    \label{fig:grad_cam}
\end{figure*}

We use two streams to learn synthesizer features and synthesizer-independent content features, respectively. To verify the effectiveness of this feature decomposition strategy, we build a single-stream network by discarding the \review{synthesizer stream, components in content streams,} and final feature fusion module. We train the single-stream ablation network using only the classification loss and the contrastive loss:
\begin{equation}
    \begin{aligned}
        \mathcal{L} = \mathcal{L}_{cls} + 0.5 * \mathcal{L}_{con\_{cls}}.
    \end{aligned}
\end{equation}
The fifth setting \review{``Single-Stream"} of Table~\ref{tab:betas} shows the detection performance of this single-stream network. One can see that the detection performance downgrades heavily on the test tasks. Without the feature decomposition strategy, the detection performance drops by about $6\%$ in the two ablation tasks. \review{Additionally, we conduct two additional ablation experiments that separately remove the content stream and the synthesizer stream. As can be seen from the settings ``No synthesizer stream" and ``No content stream" in Table~\ref{tab:betas}, removing either stream results in a degradation of detection performance.}
This performance degradation confirms our viewpoint that synthesizer-independent content features are critical for detection generalization.

To better demonstrate the effectiveness of our two-stream learning, we use the gradient-weighted class activation mapping (Grad-CAM)~\cite{selvaraju2017grad-cam} technique to visualize the gradients of the two branches. The Grad-CAM visualization can clearly show the regions focused on by the two branches. We select the model trained on the LibriseVoc cross-method task and randomly select five fake \review{speech samples} from the DiffWave subset for visualization. The visualization results are shown in Fig.~\ref{fig:grad_cam}, where we list the raw log-scale spectrogram and the heatmaps generated from the synthesizer and content streams. As can be seen, these two streams have different regions of interest. However, the content stream focuses on larger regions than the synthesizer stream, especially at time dimensions. This is because we employ two pseudo-labeling-based supervise tasks to allow the content stream to learn synthesizer-independent features, i.e., speed and compression features. These synthesizer-independent features are more coherent in the time dimension, thus enabling our content stream to be more concerned with continuity in the time dimension.

\subsection{Synthesizer Feature Augmentation}
\begin{table}[]
    \centering
        \caption{Ablation results (EER (\%)) of the feature augmentation strategy and the adversarial learning.}
    \label{tab:ablation_feature_augmentation}
    \begin{tabular}{c|ccccccc}
    \toprule
    Setting & \tabincell{c}{Feature\\Shuffle} & \tabincell{c}{Feature\\Blending} & \tabincell{c}{\review{$\mathcal{L}_{cls\_s}^*$}} &  Task1 & Task2 \\
    \midrule
    (a) & \ding{55} & \ding{55} & \ding{52} & \review{7.06} & \review{4.69} \\
    (b) &  \ding{55} & \ding{52}& \ding{52} & \review{7.09} & \review{4.34} \\
    (c) & \ding{52} & \ding{55}  & \ding{52}& \review{7.03} & \review{2.59} \\
    (d) & \ding{52} & \ding{52} & \ding{55} & \review{7.47} & \review{2.60} \\
    (e) & \ding{52} & \ding{52} & \ding{52} & \review{\textbf{6.12}} & \review{\textbf{2.18}} \\
     \bottomrule
    \end{tabular}
\end{table}

We design a synthesizer feature augmentation strategy to improve the robustness of the final classifier to different synthesizer characteristics. This strategy consists of two operations: feature shuffle and feature blending. We discard these two operations for training and testing to verify their effectiveness and provide the ablation results in Table~\ref{tab:ablation_feature_augmentation}. By comparing the test results of \review{the default setting} to those of settings (a-c), it is clear that our method using feature shuffle and feature blending together gets the best EER performance on the two ablation tasks. These results demonstrate the effectiveness of the synthesizer feature augmentation strategy on the generalization ability of our method.

\subsection{Adversarial Learning}
In the content stream, we employ adversarial learning to suppress the detection accuracy of the synthesizer based on the learned content features, i.e., maximally eliminate the synthesizer-related components in the content features. To demonstrate its effectiveness in model training, we discard it from the loss function and test the model performance. As can be seen from the settings (d-e) in Table~\ref{tab:ablation_feature_augmentation}, the adversarial learning brings about \review{$0.4\% \sim 1.35\%$} EER improvements on the two ablation tasks, which proves its necessity in the content feature learning.

\subsection{Contrastive Loss}

\begin{figure}[t]
    \centering
        \begin{minipage}[b]{0.8\linewidth}
            \centerline{\includegraphics[width=1\linewidth]{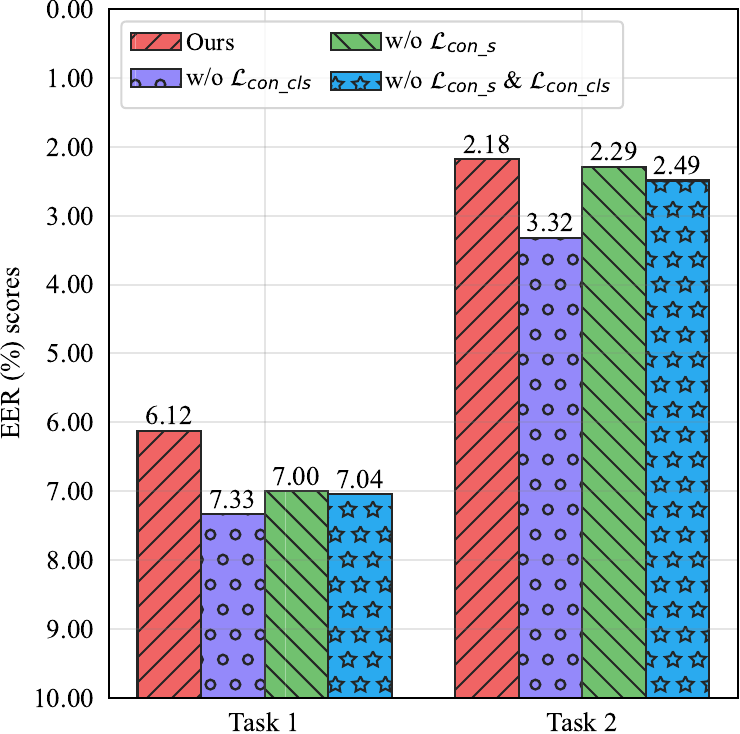}}
        \end{minipage}
    \caption{Ablation results (EER (\%)) of the contrastive losses on two ablation tasks.}
    \label{fig:hist_constrastive_loss}
\end{figure}

The contrastive loss encourages similar sample pairs to have similar representations while pushing dissimilar sample pairs apart. We utilize two contrastive losses $\mathcal{L}_{con\_s}$ and $\mathcal{L}_{con\_{cls}}$  to enhance the discriminability of the synthesizer feature $\mathbf{F}_s$  and the latent feature $\mathbf{F}_{cls}$, respectively. We apply ablation studies on both losses, and Fig.~\ref{fig:hist_constrastive_loss} illustrates the ablation results. As can be seen, using these two losses can bring about 1\% improvements in EER scores, showing the importance of contrastive loss in feature learning.

\subsection{Parameter Setting of Loss Function}

There are some hyper-parameters \review{($\beta_0, \beta_1, \beta_2, \beta_3$)} in our loss function as shown in Eq.~\eqref{eq:loss_function}. Since finding the optimal settings through an exhaustive search will be hugely time-consuming, we empirically conduct several parameter combinations to find relatively better parameter settings. Specifically, we test four parameter combinations in addition to the default setting, and Table~\ref{tab:betas} shows test results on the ablation tasks. The ablation results illustrate that our method is sensitive to these hyper-parameters. To get a relatively better performance, we get the default setting, as illustrated in the section of implementation details. In the future, it is possible to get better results by searching a larger range of parameter spaces.

\subsection{\review{Objective of Content Stream}}

\begin{table}[t]
    \centering
    \small
    \color{TableColor}
    \caption{\review{Ablation results (EER (\%)) of loss functions on two tasks.}}
    \label{tab:side_learning}
    \begin{tabular}{lrr}
    \toprule
    Setting of content stream  & Task1 & Task2 \\
    \midrule
     No pseudo-labeling-based losses  & 12.52 & 10.25 \\
     F0 prediction & 12.26 & 7.04 \\
     Default & \textbf{6.12} & \textbf{2.18}\\ 
     \bottomrule
    \end{tabular}
\end{table}

\review{We choose speech speed and compression predictions as the pseudo-labeling-based tasks for our content stream. They involve compressing and changing the speed of the input speech, which increases data diversity and enhances model generalizability. To further demonstrate the effectiveness of our method, we change the content stream's objective from speech speed and compression predictions to the prediction of the fundamental frequency (F0)~\cite{cuccovillo_audio_2023_f0}. The ablation results are shown in Table~\ref{tab:side_learning}. As can be seen, incorporating any side-learning tasks improves model performance compared to not using pseudo-labeling-based losses. Our default method can obtain better performance on the two ablation tasks, indicating the effectiveness of the chosen objectives. 
}

\subsection{\review{Real-World Application}}

\review{We method can effectively address the growing threat of malicious deepfake speech. Specifically, it can detect deepfake speech that mimics trusted individuals, avoiding fraud, financial scams, or identity theft. Another application is protecting media and communication platforms, such as social media and online conferencing tools, by integrating our detection system to identify and flag potentially harmful or misleading audio content. Additionally, our method can enhance security in voice-based authentication systems.}

\section{Conclusion}
\label{sec:conclusion}
This work proposed a robust deepfake speech detection method using feature decomposition learning and synthesizer feature augmentation. Our method aims to learn synthesizer-independent features as complementary for detection. We first designed the feature decomposition strategy that decomposes the audio representation into the synthesizer-independent content feature and synthesizer feature. The final detection is done by fusing these two kinds of features. Then, we proposed the pseudo-labeling-based supervised learning method in the content stream to learn content features and employed adversarial learning to reduce the synthesizer-related components in the content features. Finally, we introduced a synthesizer feature augmentation strategy to improve the model's robustness further. Experimental results on three benchmark deepfake speech datasets demonstrated the superior performance and generalization ability of our method.

Future research includes employing self-supervised learning tasks to learn content features rather than using pseudo-labeling-based supervised tasks. \review{Pseudo-labeling-based supervised tasks may have limited guidance for feature learning on some datasets if the speech signals have already been compressed or altered without corresponding labels.} Self-supervised learning tasks, such as masking and predicting speech embeddings, can guide the model in understanding more generalized content features. Such training requires more training data and skills. We will explore suitable ways to employ self-supervised learning tasks in the content stream.

\bibliographystyle{IEEEtran}
\bibliography{main}

\end{document}